\documentclass[aps,10pt,prd,notitlepage,tightenlines,nofootinbib,twocolumn,superscriptaddress]{revtex4-2}

\usepackage{amsmath}
\usepackage{amssymb,amsthm}
\usepackage{mathtools}
\usepackage{mathrsfs}
\usepackage{relsize}
\usepackage{bm}
\usepackage{braket}
\usepackage{booktabs}

\usepackage{epsfig,graphics,graphicx,color,xcolor}
\usepackage{soul} 
\usepackage{cancel} 
\usepackage{slashed}	
\usepackage{subfigure}
\usepackage{array}
\usepackage{multirow}
\usepackage{ulem}

\usepackage{ragged2e}
\usepackage{lineno}
\usepackage{natbib}
\usepackage{hyperref}
\hypersetup{colorlinks=true,linkcolor=red,anchorcolor=red,citecolor=orange, filecolor=brown,urlcolor=red,bookmarksnumbered=true,
pdfview=FitB
}
\graphicspath{ {image/} }
\usepackage{enumitem}

\colorlet{darkgreen}{green!50!black}
\colorlet{brightyellow}{yellow!75!red}
\colorlet{orange}{red!50!yellow}
\colorlet{darkgray}{gray!50!black}

\def\dd{{\mathrm{d}}}

\makeatletter
\newcommand*{\transpose}{%
  {\mathpalette\@transpose{}}%
}
\newcommand*{\@transpose}[2]{%
  \raisebox{\depth}{$\m@th#1\intercal$}%
}
\makeatother

\begin{document}

\title{Quantum entanglement within quarkonium}

\author{Wenyu Zhang}
\affiliation{Department of Modern Physics, University of Science and Technology of China, Hefei, Anhui 230026, China}

\author{Yiyu~Zhou}
\affiliation{Institute of Modern Physics, Chinese Academy of Sciences, Lanzhou, Gansu 730000, China}
\affiliation{University of Chinese Academy of Sciences, Beijing 100049, China}

\author{Yang Li}
\thanks{Corresponding author}
\affiliation{Department of Modern Physics, University of Science and Technology of China, Hefei, Anhui 230026, China}
\affiliation{Anhui Center for Fundamental Sciences in Theoretical Physics, Hefei, 230026, China}

\author{Qun Wang}
\affiliation{Department of Modern Physics, University of Science and Technology of China, Hefei, Anhui 230026, China}
\affiliation{Anhui Center for Fundamental Sciences in Theoretical Physics, Hefei, 230026, China}
\affiliation{School of Mechanics and Physics, Anhui University of Science and Technology, Huainan, Anhui 232001, China}

\date{\today}

\begin{abstract}
We investigate quark-antiquark entanglement in heavy quarkonium within a nonperturbative light-front Hamiltonian framework. By tracing over the antiquark degrees of freedom in the hadronic state vector, we construct the reduced density matrix of the quark subsystem and compute the associated von Neumann entropy. For spin-0 quarkonia, we show that this entropy reduces to the Shannon entropy of the unpolarized transverse momentum dependent parton distribution (TMD), up to constant color and spin contributions. For spin-1 quarkonia, we derive the explicit polarization dependence of the entropy and connect it to polarized and tensor-polarized TMDs. Using light-front wave functions obtained via basis light-front quantization (BLFQ), we evaluate the entanglement entropy for charmonium and bottomonium states, revealing a pronounced sensitivity to the polarization of vector mesons. Furthermore, we resolve the infrared parameter by matching the momentum-space entropy to a harmonic-oscillator representation. Ultimately, these results establish entanglement entropy as a novel probe of nonperturbative quarkonium structure, forging a direct link between quantum information measures and partonic observables.
\end{abstract}

\maketitle

\section{Introduction} \label{sec:introduction}

The quest to understand the internal structure of hadrons remains one of the most profound challenges in modern nuclear and particle physics. As the building blocks of the visible universe, hadrons, such as protons and neutrons, owe their existence to the intricate dynamics of QCD. Despite decades of research, the fundamental origins of hadronic properties, including color confinement, the emergence of hadronic mass, and the decomposition of nucleon spin, remain only partially understood. These phenomena are deeply rooted in the non-perturbative regime of QCD~\cite{Gross:2022hyw}, where the strong coupling constant becomes large and the vacuum structure of the theory dominates.

While traditional approaches frequently treat the parton distributions as probabilistic densities~\cite{ParticleDataGroup:2024cfk}, the underlying quantum nature of these constituents suggests that a more fundamental framework is required to capture the complete picture of hadronic wave functions.
In recent years, quantum information science (QIS) has emerged as a transformative toolkit for probing the structure of matter. By treating the vacuum and bound states of QCD as highly entangled quantum systems, QIS provides novel observables that go beyond traditional cross sections. A striking example is the recently established connection between final-state entanglement entropy and the elastic scattering cross section, which appears to satisfy a generalized area law in the context of particle interactions~\cite{Low:2024hvn,Low:2024mrk}. Furthermore, the Maximum Entropy Principle has proven effective in describing the dense gluon regime at small-$x$, suggesting that the saturation of gluon densities is driven by the maximization of entanglement between partons~\cite{Kharzeev:2017qzs,Kharzeev:2021yyf}. 

In the context of bound states, quantum entanglement offers a unique perspective on the lack of information inherent in the partonic picture~\cite{Zhang:2025ean}. When we probe a single parton, we effectively trace over the remaining degrees of freedom in the hadron, resulting in an entanglement entropy that can be related to the Shannon entropy of transverse momentum dependent distributions (TMDs). These TMDs, both spin-averaged and polarized (such as the Sivers or Boer-Mulders functions), encode the three-dimensional momentum structure and spin-orbit correlations of partons. 

Entanglement in quantum field theories (QFTs) has been extensively studied, most prominently through the entanglement entropy between spatial regions governed by the celebrated area law \cite{Nishioka:2018khk, Witten:2018zxz}. However, this notion of geometric entanglement is not directly tailored to collider physics, where observables are associated with particles and their correlations rather than with field configurations restricted to spatial subregions. Complementary approaches have explored entanglement in partonic structure using phenomenological models \cite{Liu:2018gae, CarrascoMillan:2018ufj, Beane:2019loz, Feal:2020myr, Liu:2022ohy, Liu:2022hto, Liu:2022qqf, Ehlers:2022oke, Benito-Calvino:2022kqa, Liu:2023zno, Grieninger:2023ufa, Florio:2023mzk, Ozzello:2025tfu, Florio:2025xup, Galvez-Viruet:2025rmy}, particularly in the valence-dominated large-$x$ regime \cite{Dumitru:2022tud, Dumitru:2023fih, Dumitru:2023qee, Dosch:2023bxj, Qian:2024fqf, Dumitru:2025bib, Kolbusz:2025vhh}, as well as in the dense gluon-dominated small-$x$ regime \cite{Hagiwara:2017uaz, Kovner:2018rbf, Peschanski:2019yah, Armesto:2019mna, Tu:2019ouv, Duan:2020jkz, Gotsman:2020bjc, Ramos:2020kaj, Kharzeev:2021yyf, Zhang:2021hra, Duan:2023zke, Gursoy:2023hge, Datta:2024hpn, Levin:2024wtl, Ramos:2025tge, Grieninger:2025wxg, Kutak:2025awi, Sheikhi:2025hep}, often invoking principles such as maximum entropy \cite{Cervera-Lierta:2017tdt, Beane:2018oxh, Hentschinski:2022rsa, Asadi:2022vbl, Asadi:2023bat, Hentschinski:2023izh, Hatta:2024lbw}. Despite the insights provided by these studies, a direct connection to the nonperturbative formalism of QFT is often absent. In particular, a first-principles derivation of partonic entanglement from hadronic wave functions remains largely unexplored.

To address this gap, we extend our previous framework~{\cite{Zhang:2025ean}} to QCD bound states by explicitly incorporating spin degrees of freedom. While our prior study established a connection between TMDs and entanglement entropy, it did not resolve the spin structure. A central motivation of the present work is to reveal how spin polarization shapes the entanglement structure and modifies the relationship between partonic entanglement and TMDs. We focus on heavy quarkonium systems, which provide a clean and controlled environment for investigating nonperturbative dynamics. Using a first-principles approach based on light-front hadronic wave functions, we derive the entanglement entropy between the constituent quark and antiquark, capturing its full dependence on the meson's spin and polarization.

This light-front approach offers several distinct advantages for elucidating partonic entanglement. First, light-front quantization provides a rigorous field-theoretic definition of partons that naturally generalizes the collinear partons found in standard factorization formulas~{\cite{Li:2014kfa,Karmanov:2016yzu,Hiller:2016itl,Brodsky:1997de,Carbonell:1998rj, Bakker:2013cea,Brodsky:2022fqy}}. Second, the light-front Hamiltonian formalism preserves a Galilean subgroup of the Poincaré group, ensuring a precise separation of center-of-mass (c.m.) and intrinsic dynamics that facilitates analytic expressions for the entanglement entropy~{\cite{Heinzl:2000ht, Miller:2000kv, Vary:2009gt}}. Finally, the relative simplicity of the light-front vacuum, while nontrivial in general, significantly reduces the quantum resources required for simulation~{\cite{Alterman:2025prb}}.

The remainder of the article is organized as follows. Section~\ref{sec:formalism} presents the general formalism for quantifying the entanglement entropy between quark and anti-quark in quarkonium systems, starting from the light-front wave functions (LFWFs). Section~\ref{sec:Numerical results} provides the numerical results obtained from the basis light-front quantization (BLFQ) framework, including the determination of the infrared parameter and the polarization-dependence analysis. Finally, we summarize our findings and discuss future directions in Section~\ref{sec:Summary and outlook}.

\section{Formalism}\label{sec:formalism}

\subsection{Density matrix}
To quantify the quantum entanglement among distinct constituent species within a hadron, we decompose the total Hilbert space into three subspaces, each labeled by a particle species,
\begin{equation}
    \mathcal{H}=\mathcal{H}_q\otimes\mathcal{H}_{\bar{q}}\otimes\mathcal{H}_g,
\end{equation}
where $\mathcal{H}_q(\mathcal{H}_{\bar{q}})$ is the Hilbert space containing only quarks (anti-quarks), $\mathcal{H}_{g}$ is the Hilbert space of gluons.

For a pure bipartite state $\ket{\Psi}_{AB}$, the canonical measure of entanglement is the entanglement entropy, defined as the von Neumann entropy $S_{\text{vN}}$ of the reduced density matrix of either subsystem,
\begin{equation}
    S_{A}\equiv S_{\text{vN}}(\rho_A)=-\text{Tr}_A(\rho_A\operatorname{log}\rho_A),
\end{equation}
where,
\begin{equation}
    \rho_A=\text{Tr}_B(\ket{\Psi}_{AB}\bra{\Psi}_{AB}).
\end{equation}
By the Schmidt decomposition, $\ket{\Psi}_{AB}=\sum_n\sqrt{\lambda_n}\ket{n}_A\otimes\ket{n}_B$, with $\lambda_n\geq0$ and $\sum_n\lambda_n=1$, the entanglement entropy becomes,
\begin{equation}
    S_A=-\sum_n\lambda_n\ \text{log}\ \lambda_n,
\end{equation}
which vanishes if and only if the state is a product state (i.e., only one $\lambda_n=1$).

In QCD, the reduced density matrix for the quark subsystem is obtained by tracing out the anti-quark and gluon degrees of freedom (d.o.f.'s),
\begin{equation}
    \rho_q=\text{Tr}_{\bar{q},g}\ket{\Psi}\bra{\Psi}.
\end{equation}
The entanglement entropy quantifying the quantum correlations between the quark and the rest of the system is then given by the von Neumann entropy of $\rho_q$,
\begin{equation}
    S_q=S_{\text{vN}}(\rho_q)=-\text{Tr}\big(\rho_q\operatorname{log}\rho_q\big).
\end{equation}
Analogous reduced density matrices $\rho_{\bar{q}}$ and $\rho_{g}$, and their corresponding entanglement entropies $S_{\bar{q}}$ and $S_g$, can be defined for the anti-quark and gluon subsystems, respectively.

For quarkonium, the Fock representation of the state vector $\ket{\psi}$, which encodes the full quantum information of the hadron, reads, 
\begin{multline}
|\psi(P, J, \Lambda)\rangle = \sum_{s, \bar s} \int_0^1 \frac{\dd x}{2x(1-x)} \int \frac{\dd^2 \vec{k}_\perp}{(2\pi)^3} \\ \times \psi_{s\bar s}^{\Lambda}(x, \vec{k}_\perp) 
 \frac{1}{\sqrt{N_c}} \sum_{i=1}^{N_c} b^\dagger_{si}(p) 
 d^\dagger_{\bar si}(\bar p) \ket{0} + \cdots 
\end{multline}
Here, the ellipsis represents high Fock sector contributions, $N_c = 3$ and $P^\mu = (P^-, P^+, \vec P_\perp)$ is the 4-momentum of the bound state, where $P^+ = p^+ + \bar p^+$, $\vec P_\perp = \vec p_\perp + \vec{\bar{p}}_\perp$. 
$\vec{k}_\perp = \vec p_{\perp} - x\vec P_\perp$ is the relative transverse momentum of the quark, and $x = p^+ / P^+$ is its longitudinal momentum fraction. $\psi_{s\bar s}^{\Lambda}(x, \vec{k}_\perp)$ is the valence sector LFWF, $\Lambda$ is the light-front helicity of the quarkonium.  
$b^\dagger$ and $d^\dagger$ are the creation operators for the quark and antiquark, respectively. They are related to the quark field operator $q(x)$ as, 
\begin{multline}
q(x) = \sum_s \int \frac{\dd^3p}{(2\pi)^32p^+} \Big\{b_s(p) u_s(p) e^{-ip\cdot x} \\+ d^\dagger_s(p) v_s(p) e^{+ip\cdot x}\Big\}\Big|_{x^+=0},
\end{multline} 
where, $u, v$ are the covariant 4-component Dirac spinors. For heavy quarkonium, e.g. charmonium and bottomonium, we only retain the valence Fock sector. From the Wilsonian renormalization group perspective, this approximation is equivalent to adopting an effective theory at low-energy resolution. Phenomenological investigations, including leptonic and radiative transitions, show that this approximation is sufficient for describing states below the open flavor threshold \cite{Li:2018uif, Tang:2020org, Li:2021ejv, Wang:2023nhb}. 

From the state vector $|\Psi\rangle$, we can construct the density matrix as,
\begin{equation}
\rho = |\Psi\rangle \langle \Psi|.
\end{equation}
Here, we have adopted a normalized state vector by multiplying a wavepacket for the c.m.~motion, following our previous work \cite{Zhang:2025ean}. The use of wavepacket does not change the intrinsic hadron structures thanks to the Galilean nature of light front boosts \cite{Susskind:1967rg,Kogut:1972di}.  
The reduced density matrix for the quark is obtained by tracing out the remaining d.o.f.'s:
\begin{multline}
    \rho^{\Lambda}_q = \mathrm{Tr}_{\bar{q},g} |\Psi\rangle\langle \Psi | 
    \approx \sum_{s,s'} \int\frac{\dd^3p}{(2\pi)^32p^+} \\ \int\frac{\dd^3p'}{(2\pi)^32p'^+} \rho^\Lambda_{ss'}(p', p) \frac{1}{N_c}\sum_{i} | p', s', i \rangle \langle p, s, i |.
\end{multline}
where, $i$ is the color index and $| p, s, i \rangle = b^\dagger_{si}(p) |0\rangle$. The matrix elements is, 
\begin{equation}\label{eqn:reduced_matrix_matrix_element}
\rho^\Lambda_{ss'}(p', p) = \sum_{\bar{s}} \int \frac{\dd^3p_{\bar{q}}}{(2\pi)^32p_{\bar{q}}^+}
    \Psi^{\Lambda*}_{s'\bar{s}}(p', p_{\bar{q}}) \Psi^{\Lambda}_{s\bar{s}}(p, p_{\bar{q}}).
\end{equation}
Here, $\Psi^{\Lambda*}_{s'\bar{s}}(p', p_{\bar{q}}) = \Psi(P)\psi(x, \vec{k}_\perp)$ is the  wave function with a wavepacket $\Psi(P)$. Since the form of the wavepacket does not alter the LFWF $\psi(x, \vec{k}_\perp)$, we adopt a Gaussian wavepacket with a width $\sigma$:
\begin{equation}
    \Psi(P)=\mathcal N\sqrt{2P^+}\operatorname{exp}\Big\{-\frac{(P-P_0)^2}{2\sigma^2}\Big\},
\end{equation}
where $\mathcal N$ is a normalization constant.

The reduced density matrix $\rho^{\Lambda}_q$ is not diagonal as required in Eq.~(\ref{eqn:reduced_matrix_matrix_element}). In order to proceed, we need to diagonalize the density matrix. In Ref.~\cite{Zhang:2025ean}, we showed that the reduced density matrix becomes diagonal in the narrow wavepacket limit (NWL) $\sigma \to 0$. In this limit, the Gaussian wavepacket becomes a Dirac-$\delta$ centered around the c.m. momentum $P^\mu$, forcing a momentum conservation: $p+p_{\bar q} = p' + p_{\bar q} = P$, and $\rho^\Lambda_{ss'}(p', p)$ becomes diagonal in terms of momentum variables $p$ and $p'$. 
The reduced density matrix now becomes,
\begin{multline}
    \rho^{\Lambda}_q = \sum_{s,s'} \int \frac{\dd^3p}{(2\pi)^32p^+} \rho^\Lambda_{ss'}(x, \vec{k}_\perp) \\ \times\frac{1}{N_c}\sum_{i} |p,s',i\rangle \langle p,s,i|.
\label{reduced density matrix of quark}
\end{multline}
Here, $x=p^+/P^+, \vec{k}_\perp = \vec p_\perp - x\vec P_\perp$, and 
\begin{equation}\label{eqn:density_matrix_element}
 \rho^\Lambda_{ss'}(x, \vec{k}_\perp) =  \sum_{\bar{s}} \frac{\psi^{\Lambda*}_{s'\bar{s}}(x, \vec{k}_{\perp})\psi^{\Lambda}_{s\bar{s}}(x, \vec{k}_{\perp})}{2x(1-x)(2\pi)^3}. 
\end{equation}
The spin part of the density matrix is still not diagonal. For a fixed set of relative momenta $(x, \vec{k}_\perp)$, the quark spin density matrix can be written in a matrix form: $\big[\rho^{\Lambda}\big]_{ss'} \equiv \rho^\Lambda_{ss'}$, where, 
\begin{equation}\label{eqn:spin_density_matrix}
    \rho^{\Lambda}   
    =  \frac{1}{2}
    \begin{pmatrix}
        \Phi^{[\gamma^+]}_{\Lambda} + \Phi^{[\gamma^+\gamma_5]}_{\Lambda}  & \Phi^{[i\sigma^{1+}\gamma_5]}_{\Lambda} + i\Phi^{[i\sigma^{2+}\gamma_5]}_{\Lambda}  \\
        \Phi^{[i\sigma^{1+}\gamma_5]}_{\Lambda} - i\Phi^{[i\sigma^{2+}\gamma_5]}_{\Lambda}  & \Phi^{[\gamma^+]}_{\Lambda} - \Phi^{[\gamma^+\gamma_5]}_{\Lambda} 
    \end{pmatrix}.
\end{equation}
Here, $\Phi^{[\Gamma]}_{\Lambda}$ are the quark-quark correlator functions, which can be parameterized in terms of quark TMDs~\cite{Boussarie:2023izj},
\begin{multline}
    \Phi^{[\Gamma]}_{\Lambda}(x,\vec{k}_{\perp})=\frac{1}{2}\int\frac{\dd z^-\dd^2z_{\perp}}{2(2\pi)^3}e^{ikz}\langle\psi(P, j, \Lambda)|\\ \bar{\psi}_q(0)\Gamma\mathcal{W}(0_{\perp},z_{\perp})\psi_q(z)|\psi(P, j, \Lambda)\rangle\Big|_{z^+ = 0},
\label{quark-quark correlator function}
\end{multline}
with $k^+=xP^+$. $\mathcal W(0_\perp, z_\perp)$ is the Wilson line that maintains the gauge invariance. 
In the valence sector ansatz with approximation $\mathcal W = 1$, the quark-quark correlator functions can be represented by LFWFs as, 
\begin{multline}
    \Phi^{[\Gamma]}_{\Lambda}(x,\vec{k}_{\perp})=\frac{1}{2p_1^+}\\
    \sum_{s,r,\bar{s}}\frac{\psi^{\Lambda*}_{r\bar{s}}(x,\vec{k}_{\perp})\psi^{\Lambda}_{s\bar{s}}(x,\vec{k}_{\perp})}{2x(1-x)(2\pi)^3}
    \bar{u}_r(p_1)\Gamma u_s(p_1),
\label{meson quark TMD}
\end{multline}
which are exactly the matrix elements of the density matrix in Eq.~(\ref{eqn:density_matrix_element}).

\subsection{Spin-0}

For spin-0 mesons, the leading-twist quark-quark correlators can be parametrized in terms of TMDs as~\cite{Pasquini:2014ppa},
\begin{align}
    \Phi^{[\gamma^+]}(x, \vec{k}_\perp) =\,& f_1(x, {k}_\perp), \\
    \Phi^{[\gamma^+\gamma^5]}(x, \vec{k}_\perp) =\,& 0, \\
    \Phi^{[i\sigma^{\alpha+}\gamma^5]}(x, \vec{k}_\perp) =\,& \frac{\epsilon_\perp^{\alpha\beta}k_\perp^\beta}{M}h_1^\perp(x,\vec k_\perp).
\end{align}
Here, $f_1$ denotes the unpolarized TMD, while $h_1^\perp$ is the Boer-Mulders function, which arises from gauge-link effects. We adopt the convention $\epsilon_\perp^{12}=1$, and $M$ denotes the meson mass. In the approximation $\mathcal W=1$, the Boer-Mulders function vanishes, leaving $f_1$ as the only nonzero leading-twist TMD~\cite{Puhan:2023ekt},
\begin{equation}
f_1(x, {k}_\perp) = \Phi^{[\gamma^+]}(x, \vec{k}_\perp) =  \sum_{s,\bar{s}} \frac{\big|\psi_{s\bar{s}}(x, \vec{k}_{\perp})\big|^2}{2x(1-x)(2\pi)^3},
\end{equation}
Consequently, the spin density matrix in (\ref{eqn:spin_density_matrix}) becomes diagonal. The reduced density matrix for spin-0 reads, 
\begin{multline}\label{eqn:reduced_density_matrix_spin-0}
    \rho_q = \frac{1}{2}\sum_{s} \int \frac{\dd^3p}{(2\pi)^32p^+} f_1(x,   {k}_\perp) \\ 
    \times\frac{1}{N_c}\sum_{i} |p,s,i\rangle \langle p,s,i|, 
\end{multline}
where, $x = p^+/P^+_0$, and $\vec{k}_\perp = \vec p_\perp - x \vec P_{0\perp}$. 
The corresponding entanglement entropy is directly related to the Shannon entropy of the quark TMD $f_1$,
\begin{multline}
    S_\text{vN}(\rho_q) = \log \bigg(\frac{P^+_0VN_c}{4\pi^3}\bigg) \\
    - \int \dd x \int \dd^2   {k}_\perp \, f_1(x,  {k}_\perp) \log f_1(x,  {k}_\perp) \\
= \log(2N_c) + H(f_1).
\end{multline}
Here, $H(f_1)$ is the Shannon entropy of the unpolarized quark TMD $f_1$, while $\operatorname{log}N_c$ and $\operatorname{log}2$ represent the contributions from color entanglement and spin entanglement, respectively. $V = L_\perp^2 L$ is the spatial volume in the box regularization, which also serves as an IR parameter, with $L$ and $L_\perp$ being the box lengths in the longitudinal and transverse directions, respectively. $P_0^+L$ is a boost invariant quantity, proportional to $\delta x^{-1}$, the resolution of the longitudinal momentum~\cite{Pauli:1985ps,Pauli:1985pv}. 

\subsection{Spin-1}

For spin-1 mesons, the decomposition of the quark-quark correlator functions into TMDs is significantly more complex due to the additional degrees of freedom associated with vector and tensor polarizations. At the leading-twist level, the correlators are parameterized as follows~\cite{Bacchetta:2001rb,Ninomiya:2017ggn},
\begin{multline}
    \Phi^{[\gamma^+]}_{\Lambda}(x,\vec{k}_{\perp})=f_1(x,\vec{k}_{\perp})\\
    +S_{LL}f_{1LL}(x,\vec{k}_{\perp})+\frac{\vec{S}_{LT}\cdot\vec{k}_{\perp}}{M}f_{1LT}(x,\vec{k}_{\perp})\\
    +\frac{\vec{k}_{\perp}\cdot\boldsymbol{S}_{TT}\cdot\vec{k}_{\perp}}{M}f_{1TT}(x,\vec{k}_{\perp}),
\label{spin-1 TMD 1}
\end{multline}
\begin{multline}
    \Phi^{[\gamma^+\gamma_5]}_{\Lambda}(x,\vec{k}_{\perp})=S_{L}g_{1L}(x,\vec{k}_{\perp})\\
    +\frac{\vec{S}_{T}\cdot\vec{k}_{\perp}}{M}g_{1T}(x,\vec{k}_{\perp}),
\label{spin-1 TMD 2}
\end{multline}
\begin{multline}
    \Phi^{[i\sigma^{j+}\gamma_5]}_{\Lambda}(x,\vec{k}_{\perp})=S_{T}^jh_1(x,\vec{k}_{\perp})+S_L\frac{k_\perp^j}{M}h_{1L}^{\perp}(x,\vec{k}_{\perp})\\
    +\frac{2k_\perp^j\vec{S}_{T}\cdot \vec{k}_{\perp}-S_T^j\vec{k}_{\perp}^2}
    {2M^2}h_{1T}^{\perp}(x,\vec{k}_{\perp}).
\label{spin-1 TMD 3}
\end{multline}
In these expressions, the polarization of the meson is characterized by a spin vector $\vec{S}=(S_T^x,S_T^y,S_L)$ and a symmetric, traceless rank-2 spin tensor $T$,
\begin{equation}
    T=\frac{1}{2}
    \begin{pmatrix}
        S_{LL}+S^{xx}_{TT} & S^{xy}_{TT} & S^x_{LT} \\
        S^{yx}_{TT} & S_{LL}+S^{yy}_{TT} & S^y_{LT} \\
        S^x_{LT} & S^y_{LT} & -2S_{LL}
    \end{pmatrix}.
\end{equation}
The explicit expressions for the parameters in the polarization vectors and tensors are presented in Appendix~\ref{Appendix:polarization parameters}.
The terms $f_{1LL}$, $f_{1LT}$, and $f_{1TT}$ represent the tensor-polarized TMDs, which are unique features of spin-1 (or higher) systems and have no analogue in the spin-0 or spin-1/2 cases.

From the block diagonal form of the reduced density matrix (\ref{reduced density matrix of quark}), the entanglement entropy is,
\begin{multline}
    S_q(\Lambda)=\operatorname{log}\bigg(\frac{P_0^+VN_c}{8\pi^3}\bigg)\\-\sum_{a=1}^2\int\dd x\int\dd^2 {k}_{\perp}\lambda_a^{\Lambda}(x,\vec{k}_{\perp})\operatorname{log}\lambda_a^{\Lambda}(x,\vec{k}_{\perp}),
\label{quark entropy}
\end{multline}
where $\lambda_a^{\Lambda}\ (a=1,2)$ are the eigenvalues of the quark spin density matrix $\rho^{\Lambda}(x,\vec{k}_\perp)$,
\begin{multline}
    \lambda^{\Lambda}_{1,2}(x,\vec{k}_\perp)=\frac{1}{2}\Bigg[\Phi^{[\gamma^+]}_{\Lambda}\pm \\ \sqrt{\Big(\Phi^{[\gamma^+\gamma_5]}_{\Lambda}\Big)^2
    +\Big(\Phi^{[i\sigma^{1+}\gamma_5]}_{\Lambda}\Big)^2+\Big(\Phi^{[i\sigma^{2+}\gamma_5]}_{\Lambda}\Big)^2}\Bigg].
\label{eqn:eigenvalues of quark spin density matrix}
\end{multline}
It is noteworthy that the entanglement entropy for spin-1 mesons depends on its polarization $\Lambda$. According to Eqs.~(\ref{spin-1 TMD 1}--\ref{spin-1 TMD 3}), the entanglement entropy can be expressed in terms of the quark TMDs and the polarization parameters of the meson.
 
For example, for the eigenstate $\ket{m_J=0}$ of the angular momentum operator $\hat{J}_z$ with eigenvalue 0, the spin vector and spin tensor are,
\begin{equation}
    \vec{S}=(0,0,0),\qquad T=\operatorname{diag}\Big\{\frac{1}{3},\frac{1}{3},-\frac{2}{3}\Big\},
\end{equation}
where only $S_{LL}$ is non-zero. The eigenvalues of the quark spin density matrix become,
\begin{multline}
    \lambda_{1}^{m_J=0}(x,\vec{k}_\perp)=\lambda_{2}^{m_J=0}(x,\vec{k}_\perp)\\=\frac{1}{2}f_1(x,\vec{k}_\perp)+\frac{1}{3}f_{1LL}(x,\vec{k}_\perp),
\end{multline}
and the entanglement entropy can be simplified in terms of quark TMDs $f_1$ and $f_{1LL}$,
\begin{equation}
    S_q(m_J=0) = \log 2N_c + H\Big(f_1+\frac{2}{3}f_{1LL}\Big),
\end{equation}
where $H$ is the Shannon entropy.

For the eigenstates $\ket{m_J=\pm1}$ of the angular momentum operator $\hat{J}_z$ with eigenvalues $\pm1$, their spin vectors and spin tensors are,
\begin{equation}
    \vec{S}=(0,0,\pm1),\qquad T=\operatorname{diag}\Big\{-\frac{1}{6},-\frac{1}{6},\frac{1}{3}\Big\}.
\end{equation}
In this case, the eigenvalues of the quark spin density matrix are,
\begin{multline}
    \lambda_{1,2}^{|m_J|=1}(x,\vec{k}_\perp)=\frac{1}{2}\Bigg(f_1(x,\vec{k}_\perp)-\frac{1}{3}f_{1LL}(x,\vec{k}_\perp)\\
    \pm\sqrt{g_{1L}^2(x,\vec{k}_\perp)+\frac{\vec{k}_\perp^2}{M^2}h_{1L}^{\perp2}(x,\vec{k}_\perp)}\Bigg),
\end{multline}
which indicates that $S_q(m_J=\pm1)$ are identical, in alignment with parity conservation while the longitudinal polarization $m_J$ flips sign, $m_J\rightarrow-m_J$.

\section{Numerical results}\label{sec:Numerical results}

\subsection{LFWF from BLFQ}
In this work, we use the LFWFs obtained from a low-energy effective Hamiltonian, which incorporates the light-front kinetic energy $T$, the soft-wall AdS/QCD confining potential $V_{\text{conf}}$, and a one-gluon exchange interaction $V_{\text{OGE}}$ derived from light-front QCD \cite{Li:2017mlw},
\begin{equation}
    H_{\text{eff}}=T+V_{\text{conf}}+V_{\text{OGE}}.
\end{equation}
The hadronic state vector is obtained by solving the light-front Schrödinger equation,
\begin{equation}
    H_{\text{eff}}\ket{\psi(p)}=M^2\ket{\psi(p)},
\label{Hamiltonian eigenvalue problem}
\end{equation}
in the basis light-front quantization (BLFQ) approach~\cite{Vary:2009gt}.
In this approach, the Hamiltonian is represented as a matrix $\big[H \big]_{ij}=\braket{i|H|j}$ in a prescribed basis $\ket{i}$. The basis consists of two-dimensional harmonic oscillator functions $\phi_{nm}(\vec{q}_\perp)$ in the transverse direction and Jacobi polynomials $\chi_l(x)$ in the longitudinal direction \cite{Li:2015zda}. Here, $n,m,l$ denote the BLFQ basis quantum numbers. These bases preserve all kinematical symmetries of the QCD Hamiltonian and are also eigenfunctions of the AdS/QCD Hamiltonian $H_0 = T + V_\text{conf}$.
In practical calculations, the transverse and longitudinal bases are truncated according to their basis energies~\cite{Li:2017mlw}:
\begin{equation}
    2n+|m|+1\leq N_\text{max},\qquad 0\leq l\leq L_\text{max}.
\end{equation}
Accordingly, the $N_\text{max}$-truncation provides a natural pair of ultraviolet and infrared cutoffs,
\begin{equation}
    \Lambda_{\text{UV}}\sim\kappa\sqrt{N_{\text{max}}},\qquad \Lambda_{\text{IR}}\sim\kappa/\sqrt{N_{\text{max}}},
\end{equation}
where $\kappa$ is the basis scale parameter of the harmonic oscillator functions.
The Hamiltonian matrix is then numerically diagonalized to obtain the mass spectrum and the corresponding wave functions.
The valence-sector LFWF can be expanded in terms of the BLFQ basis as~\cite{Li:2017mlw},
\begin{multline}
    \psi^\Lambda_{s\bar{s}}(x,\vec{k}_\perp)=\sum_{n,m,l}\psi^\Lambda_h(n,m,l,s,\bar{s})\\ \times \phi_{nm}\bigg(\frac{\vec{k}_\perp}{\sqrt{x(1-x)}}\bigg)\chi_l(x),
\end{multline}
where $\psi^\Lambda_h(n,m,l,s,\bar{s})$ are the expansion coefficients obtained from diagonalization.
The resulting charmonium spectrum has been shown to be in good agreement with experimental measurements. The corresponding light-front wave functions have also been used to compute a variety of observables, including radiative widths, yielding good agreement with experimental values \cite{Li:2017mlw}.

\begin{figure*}
\centering 
\subfigure[\ ]{\includegraphics[width=.32\textwidth]{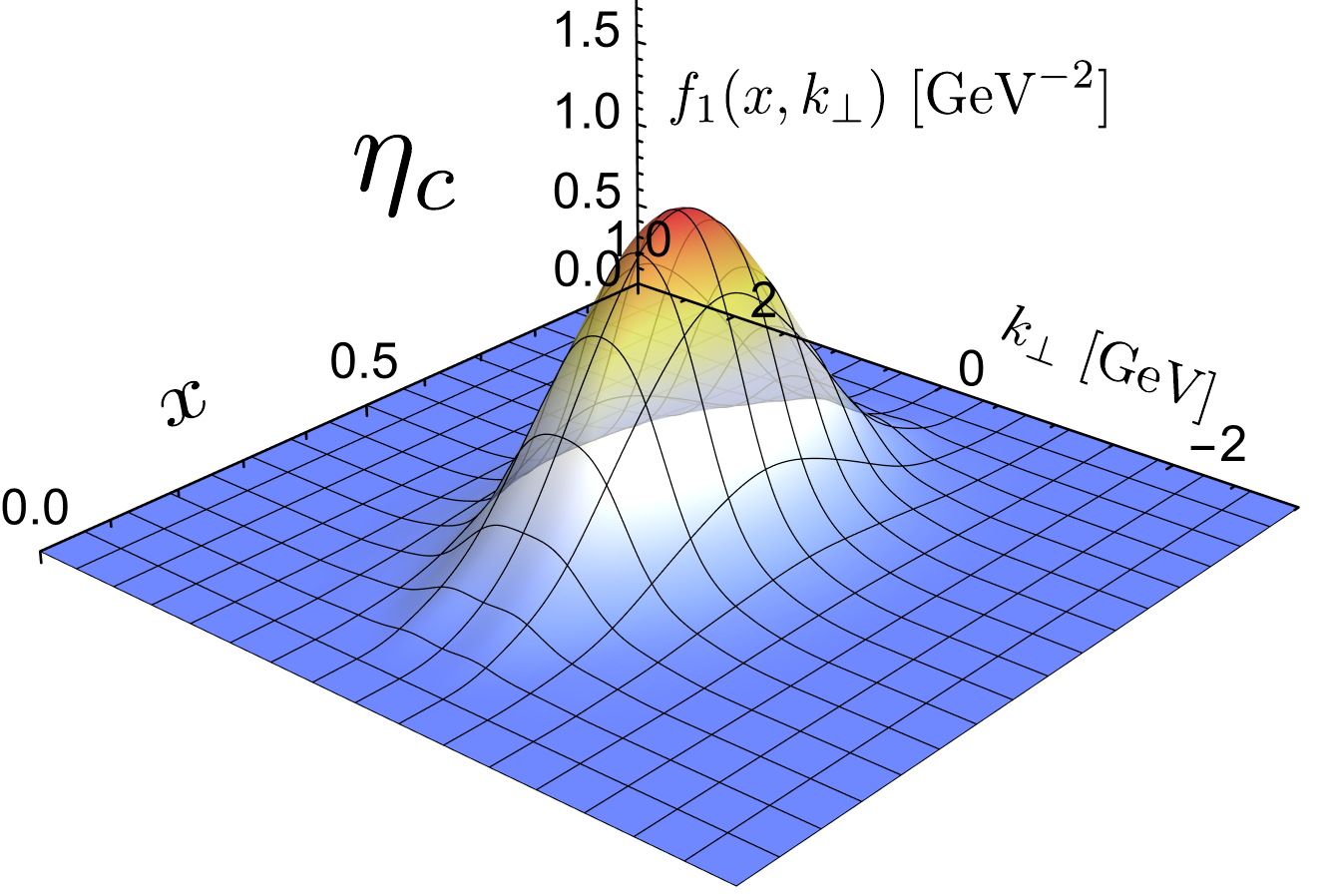}} \hfill
\subfigure[\ ]{\includegraphics[width=.32\textwidth]{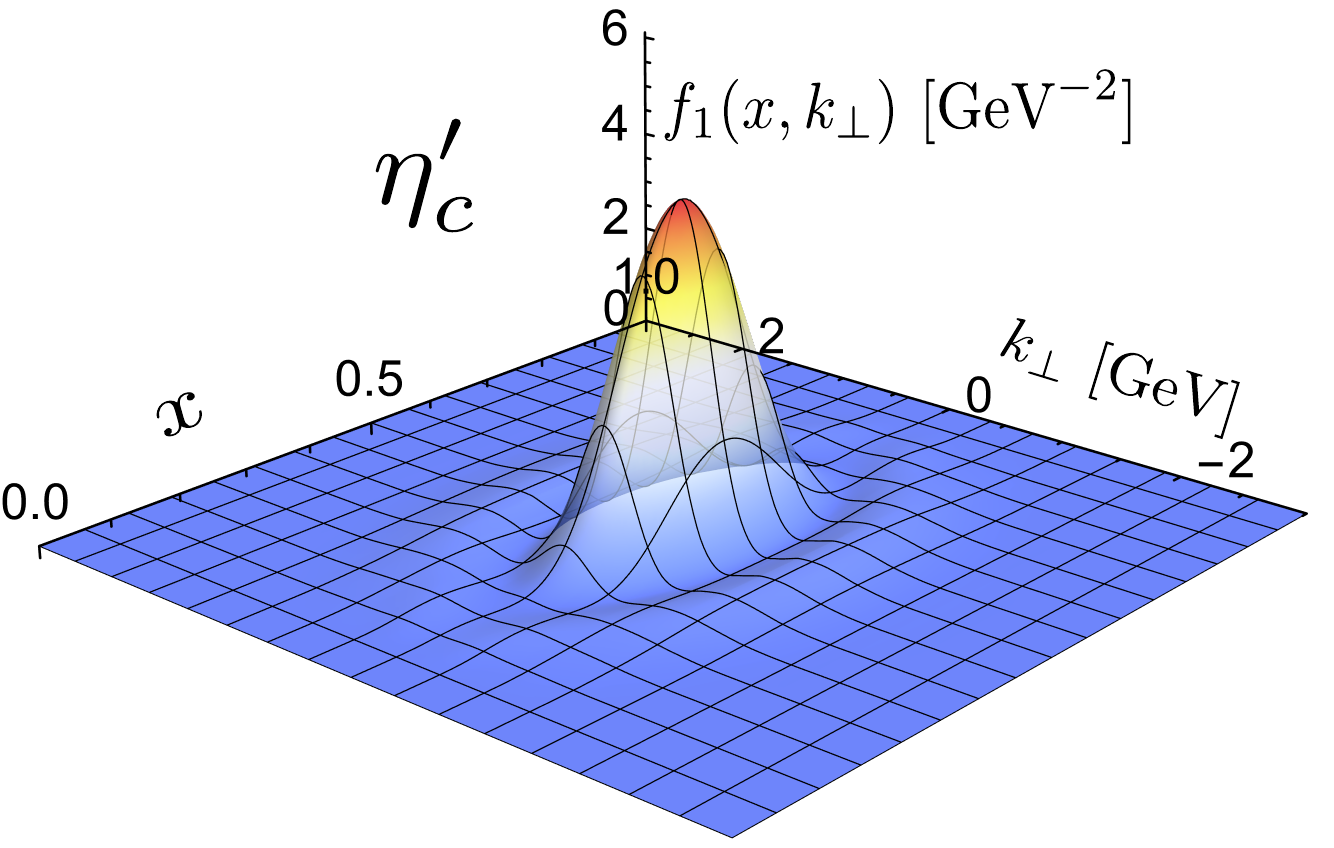}} \hfill
\subfigure[\ ]{\includegraphics[width=.32\textwidth]{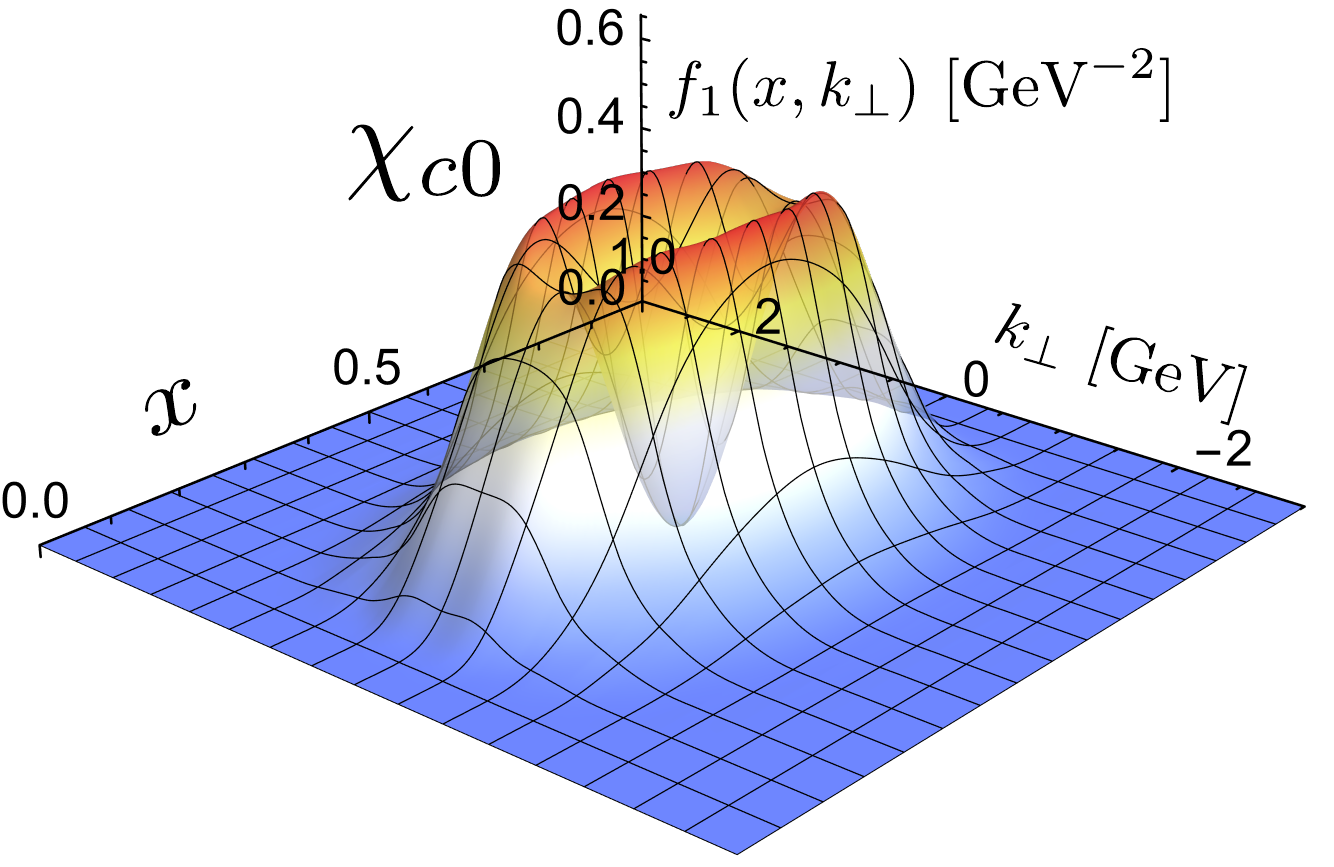}}
\caption{3D images of the unpolarized TMDs $f_1(x,  {k}_\perp)$ for $\eta_c$, $\eta_c'$ and $\chi_{c0}$.}
\label{fig:TMDs}
\end{figure*}

\begin{figure*}
\centering 
\subfigure[\ ]{\includegraphics[width=.32\textwidth]{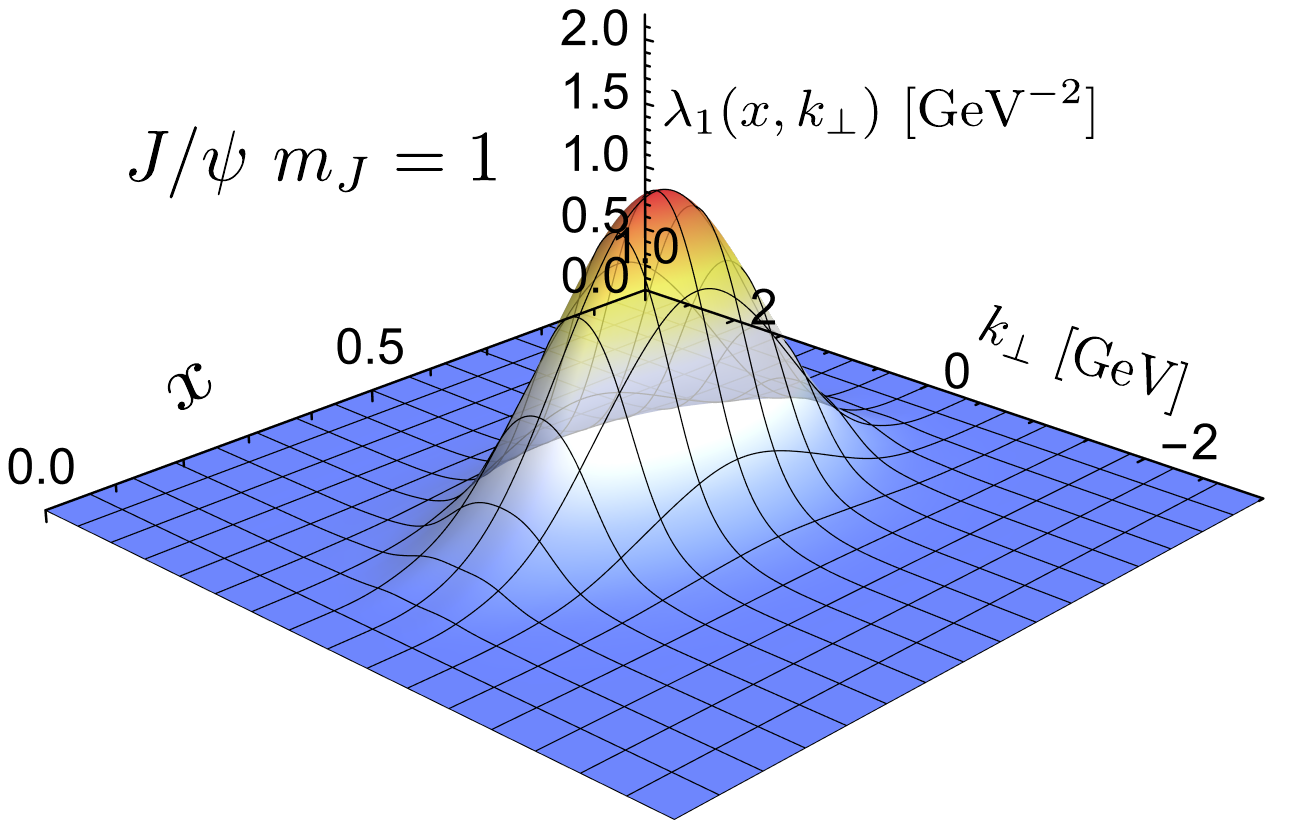}} \hfill
\subfigure[\ ]{\includegraphics[width=.32\textwidth]{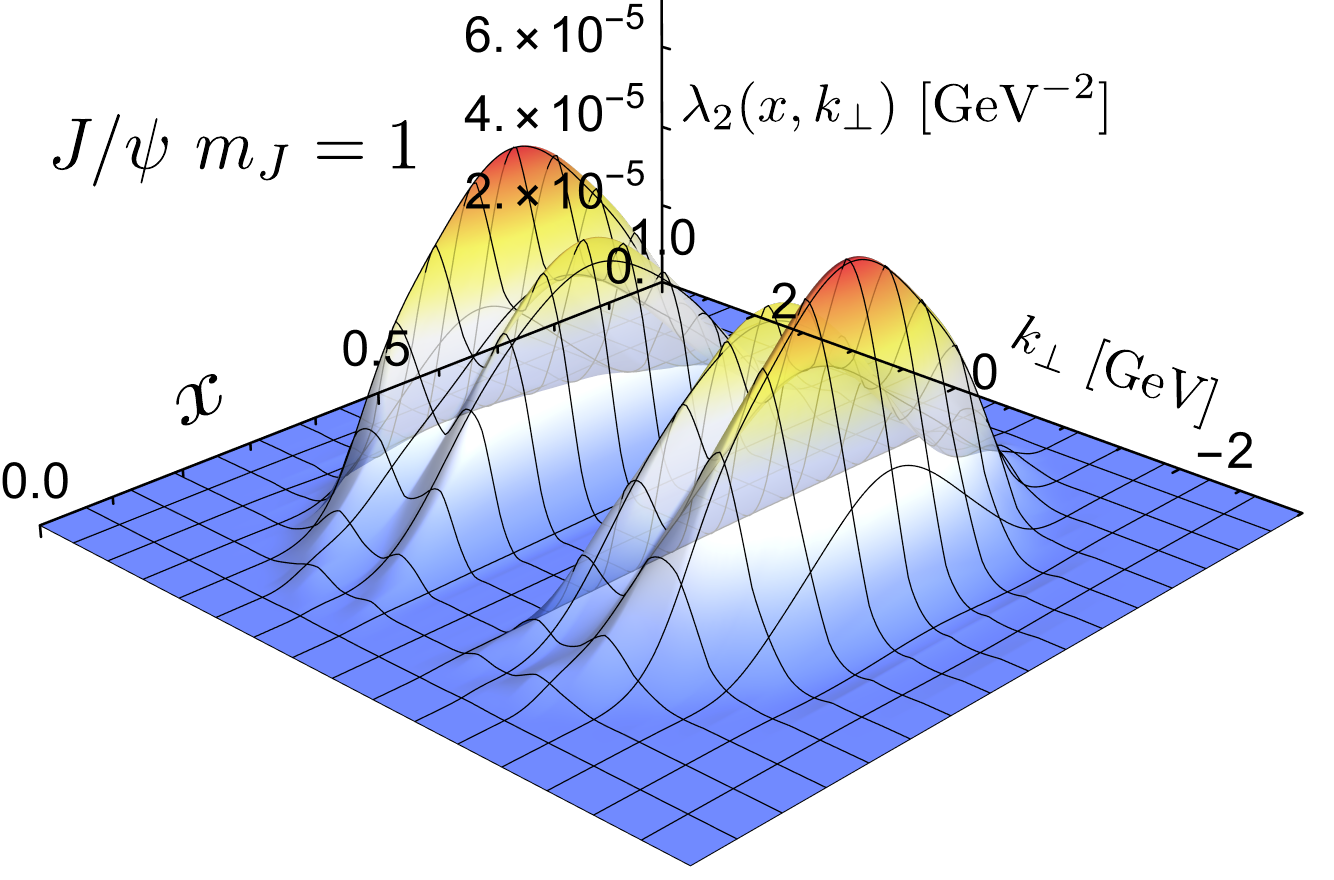}} \hfill
\subfigure[\ ]{\includegraphics[width=.32\textwidth]{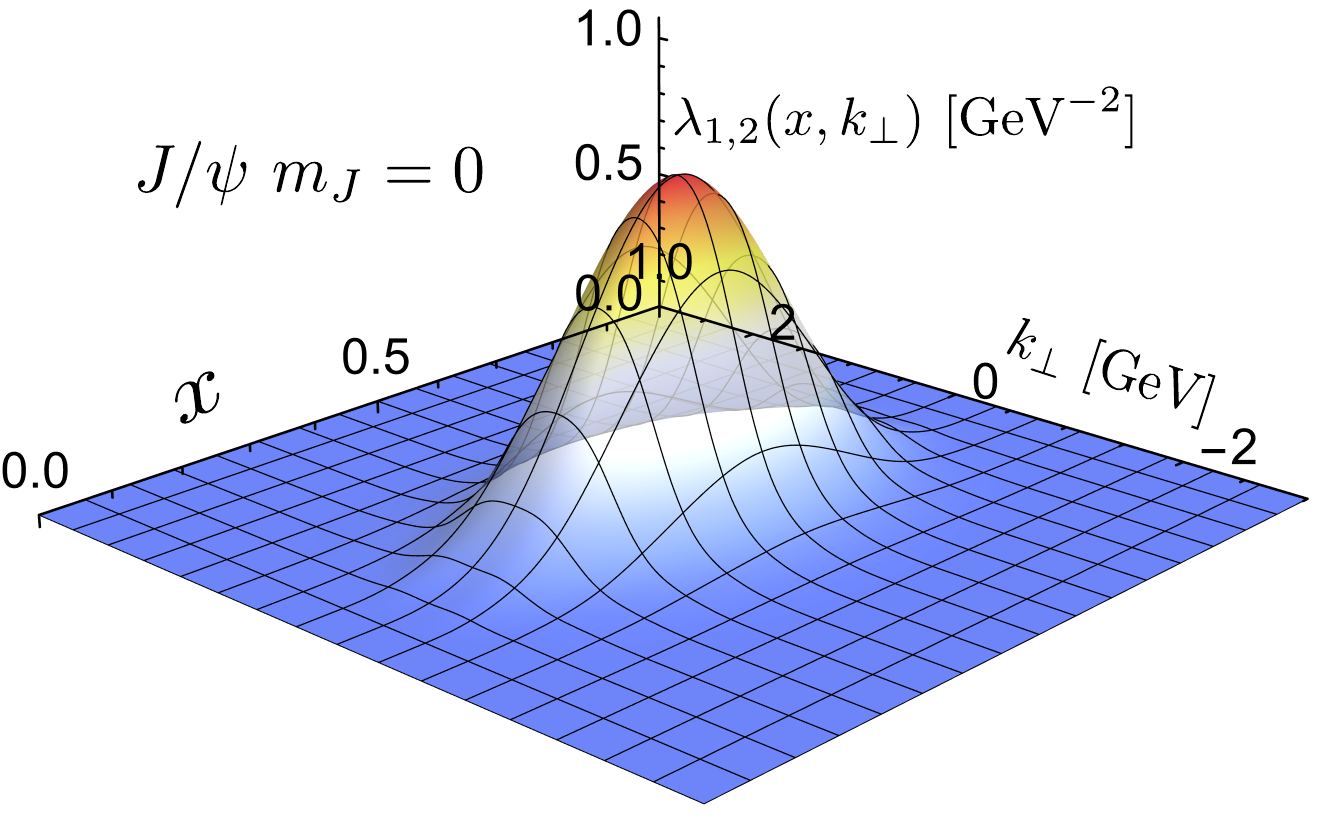}} \hfill
\subfigure[\ ]{\includegraphics[width=.32\textwidth]{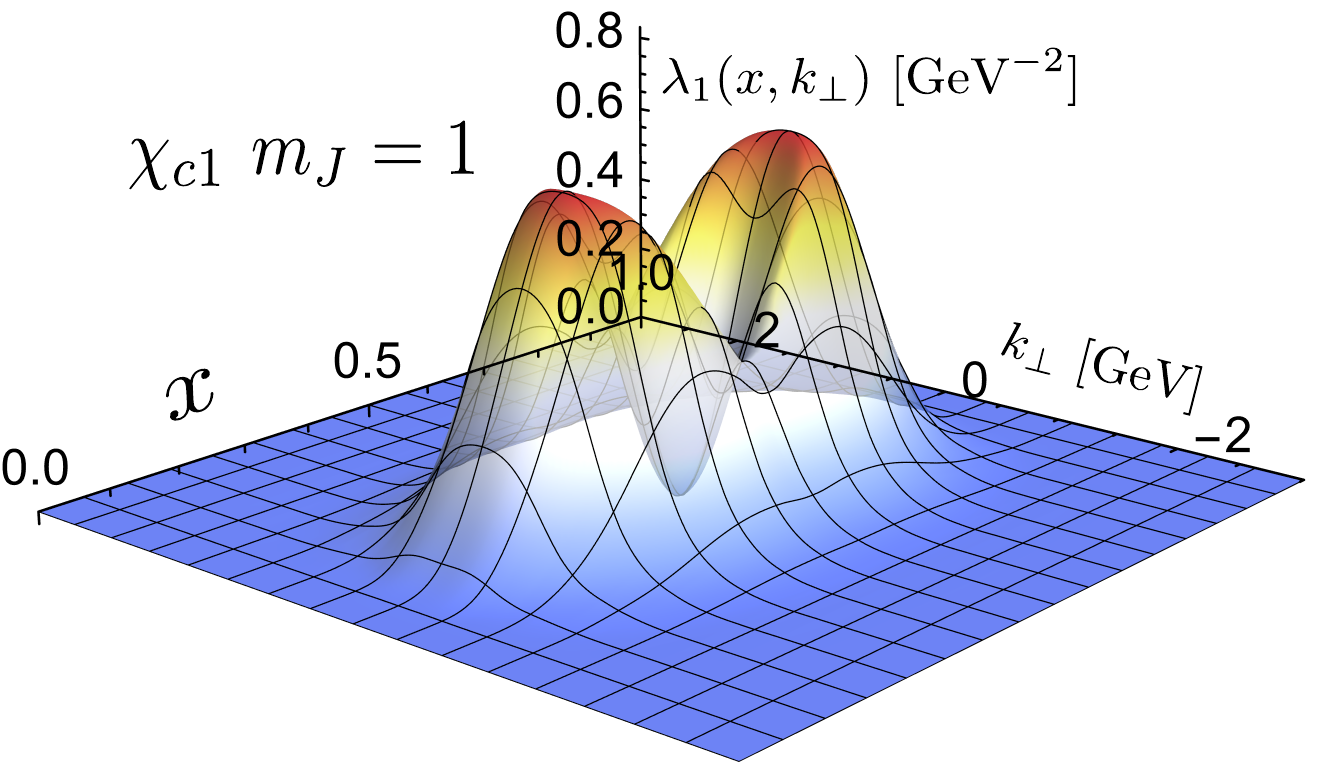}}\hfill
\subfigure[\ ]{\includegraphics[width=.32\textwidth]{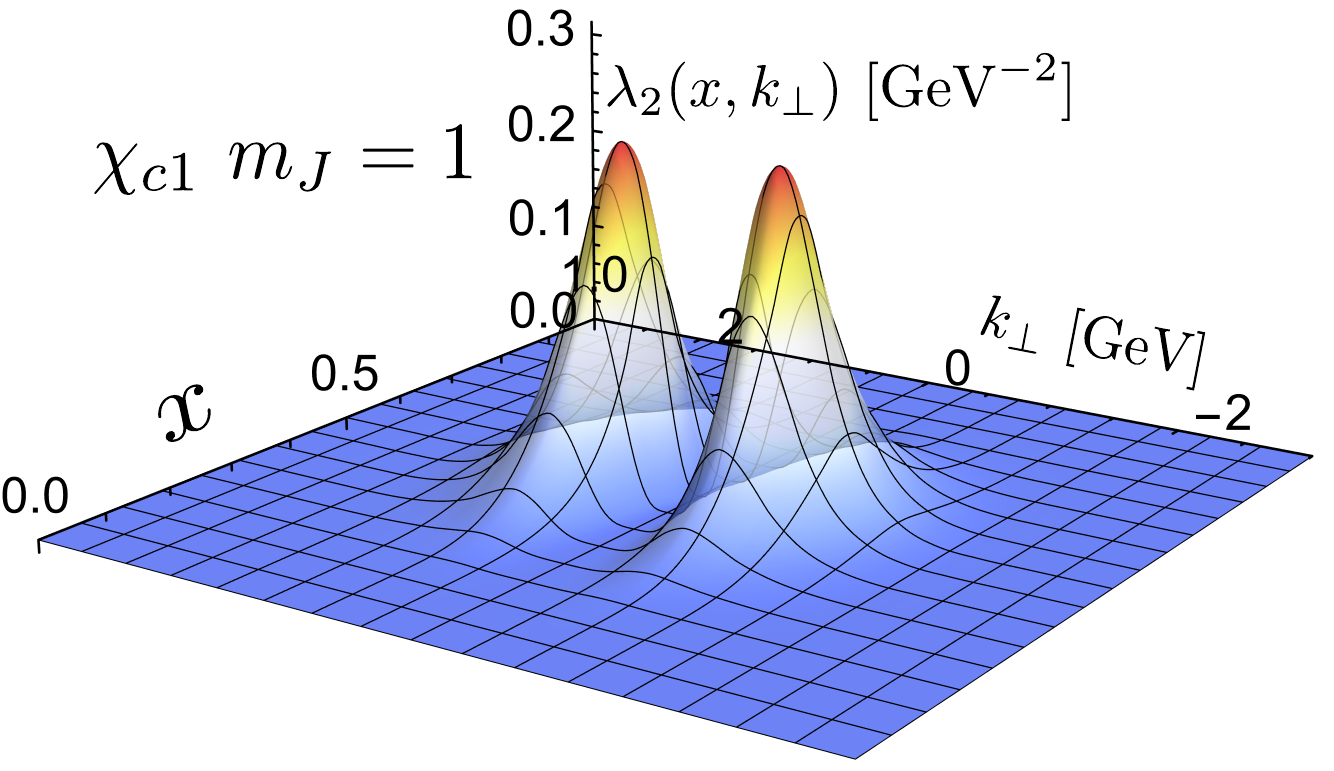}}\hfill
\subfigure[\ ]{\includegraphics[width=.32\textwidth]{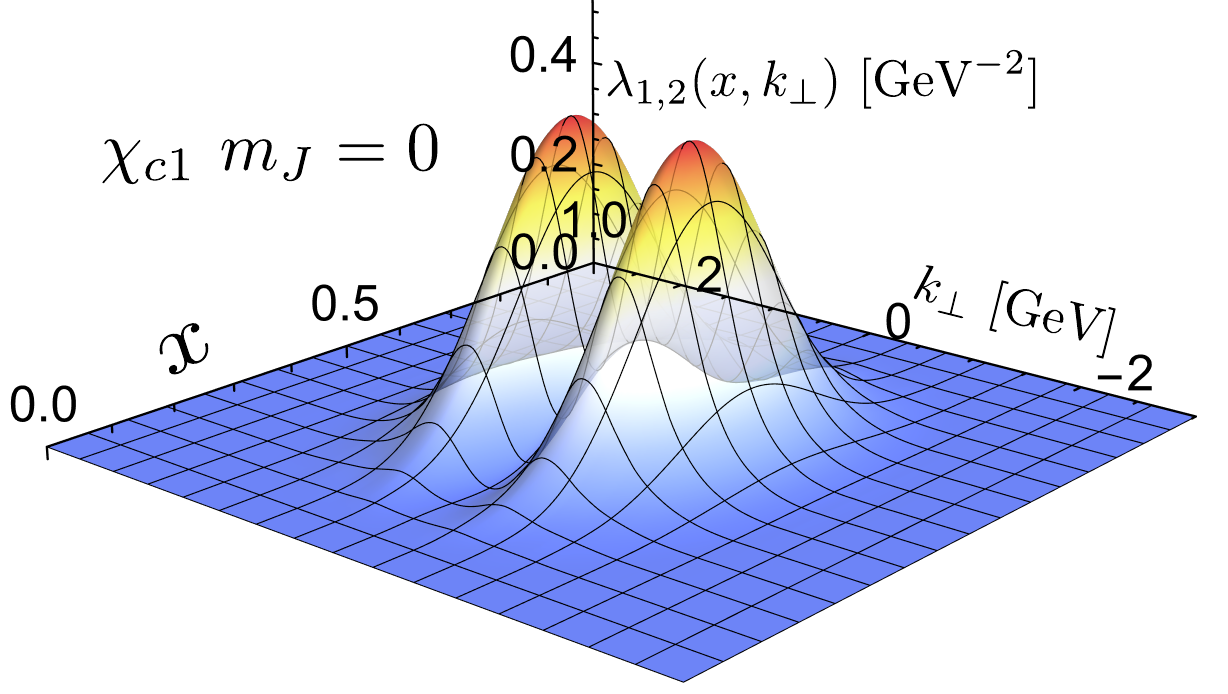}}\hfill
\subfigure[\ ]{\includegraphics[width=.32\textwidth]{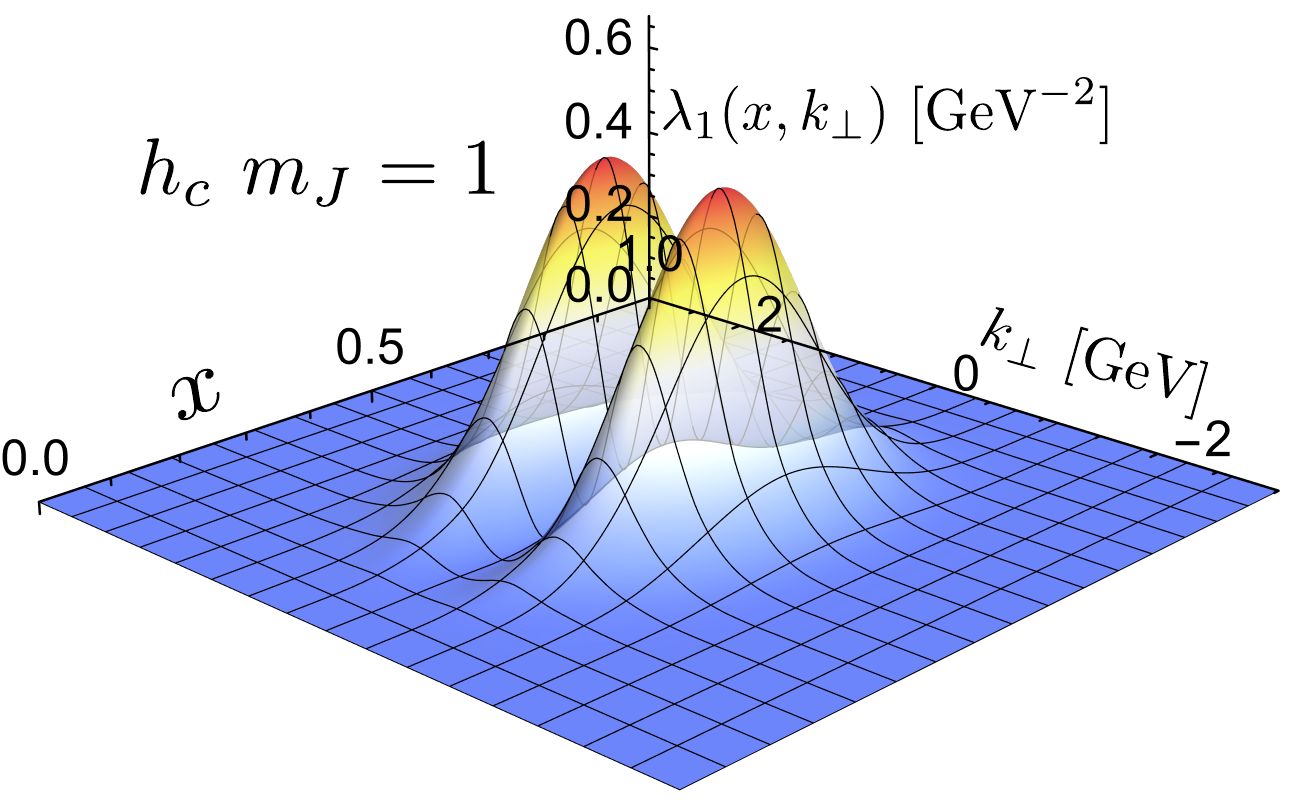}}\hfill
\subfigure[\ ]{\includegraphics[width=.32\textwidth]{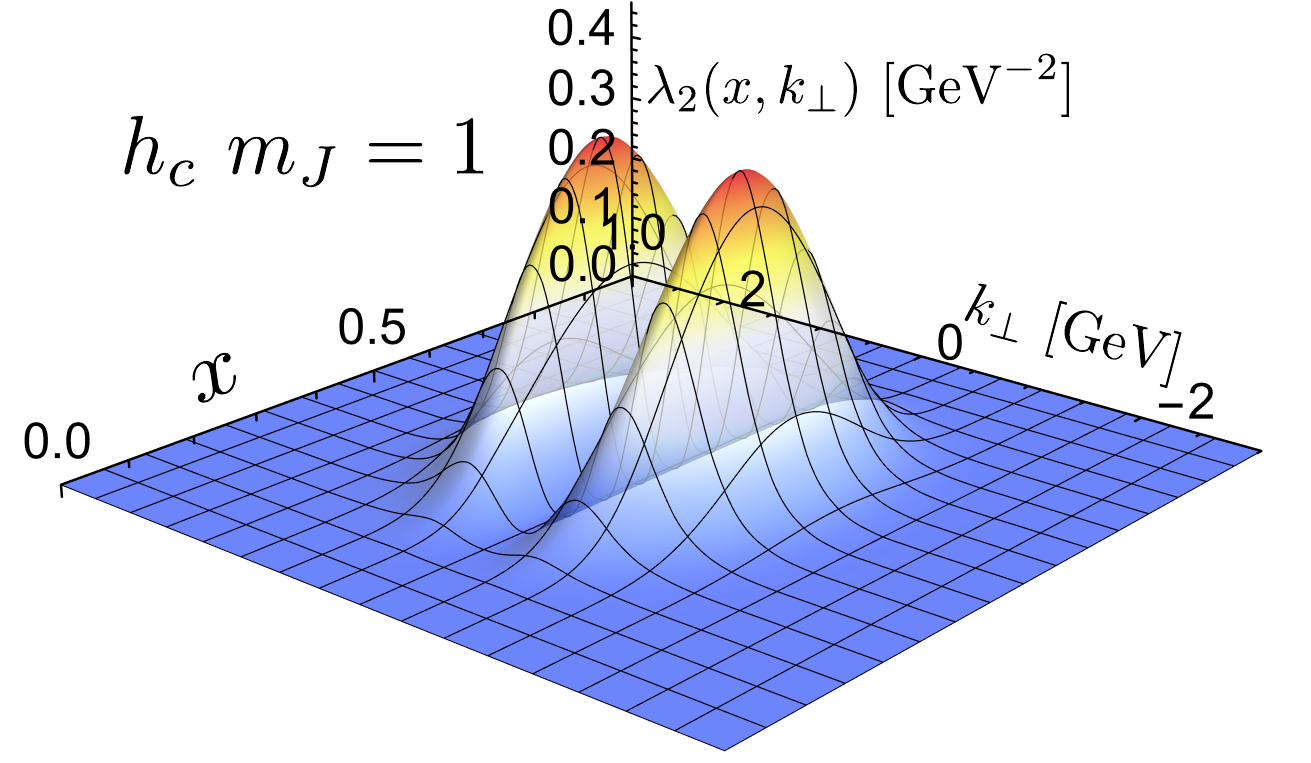}}\hfill
\subfigure[\ ]{\includegraphics[width=.32\textwidth]{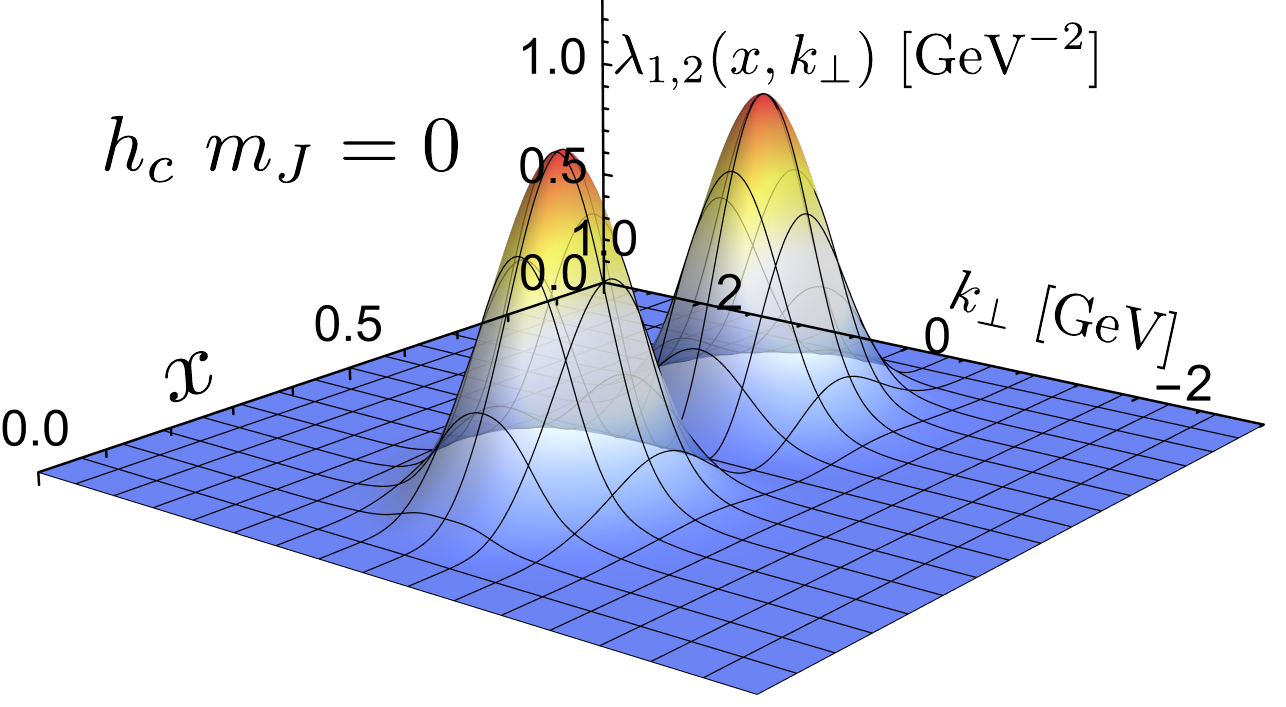}}
\caption{3D images of the eigenvalues of the quark spin density matrix $\lambda_{1,2}(x,k_\perp)$ for $J/\psi$, $\chi_{c1}$ and $h_c$.}
\label{fig:Lambdas}
\end{figure*}

Figure~\ref{fig:TMDs} illustrates the unpolarized TMDs for spin-0 charmonium ($\eta_c, \chi_{c0}, \eta_c'$) obtained within BLFQ. 
Figure~\ref{fig:Lambdas} illustrates the eigenvalues of the quark spin density matrix $\lambda_{1,2}(x,k_\perp)$ defined by Eq.(\ref{eqn:eigenvalues of quark spin density matrix}) for spin-1 charmonium ($J/\psi, \chi_{c1}, h_c$) obtained within BLFQ.

\subsection{Determination of the IR parameter}

In Section~\ref{sec:formalism}, the entanglement entropy between the quark and anti-quark degrees of freedom was evaluated in the momentum representation. In this representation, a logarithmic IR term, $\operatorname{log}(P_0^+V/(2\pi)^3)$, remains in the entropy even after implementing the UV and IR truncation schemes typical of BLFQ. To properly determine this IR parameter, the entanglement entropy can alternatively be evaluated in the harmonic oscillator representation.

The harmonic oscillator basis functions, $\phi_{mn}$, are defined in terms of the relative transverse momentum of the quark-antiquark pair. To rewrite the hadronic state in a tensor-product basis of single-particle states, the c.m.~motion must be explicitly introduced. We supplement the intrinsic wave function with a c.m.\ wave packet and impose the NWL. Adopting box regularization in the longitudinal direction (with length $L$) and a continuous Gaussian profile in the transverse direction, the c.m.\ wave packet reads,
\begin{equation}
    \Psi(P)=\sqrt{2P_0^+L}\delta_{P^+,P_0^+} \frac{\sqrt{2\pi}}{\sigma}\operatorname{exp}\Big\{-\frac{\vec{P}^2_{\perp}}{4\sigma^2}\Big\}.
\label{eqn:center-of-mass motion}
\end{equation}
After performing a Talmi-Moshinsky transformation~\cite{Li:2015iaw}, the hadronic state can be expressed as a tensor product of the BLFQ single-particle bases. Applying the appropriate mixed integration measure, we obtain
\begin{widetext}
\begin{multline}
        \frac{1}{2L}\sum_{P^+}\int\frac{\dd^2 P_\perp}{(2\pi)^2 2P^+}\Psi(P)\ket{\psi(P,J,\Lambda)}=\frac{1}{\sqrt{2P_0^+L}}\sum_{p_1^+,p_2^+}\sum_{n_1m_1n_2m_2}\sum_{s,\bar{s}}\bigg[\sum_{N,M}\sum_{n,l,m}c_{NM}\\
    \mathcal{M}_{NMnm}^{n_1m_1n_2m_2}\psi_h^\Lambda(n,m,l,s,\bar{s})\chi_l(x)\delta_{p_1^++p_2^+=P_0^+}\bigg]\ket{p_1^+,n_1,m_1,s}\otimes\ket{p_2^+,n_2,m_2,\bar{s}}.
\end{multline}
\end{widetext}
Here, $c_{NM}$ denotes the expansion coefficient of the transverse c.m.\ motion in the harmonic oscillator basis, satisfying
\begin{equation}
    \frac{\sqrt{2\pi}}{\sigma}\operatorname{exp}\Big\{-\frac{\vec{P}^2_{\perp}}{4\sigma^2}\Big\}=\sum_{N,M}c_{NM}\phi_{NM}(\vec{P}_\perp),
\end{equation}
Additionally, $\mathcal{M}_{NMnm}^{n_1m_1n_2m_2}$ represents the Talmi-Moshinsky transformation coefficient, $x=p_1^+/P_0^+$ is the longitudinal momentum fraction, and $\ket{p_i^+,n_i,m_i,s}$ denotes the orthonormal single-particle basis. Note that the momentum conservation delta in the longitudinal direction is now a discrete Kronecker delta, $\delta_{p_1^++p_2^+,P_0^+}$.

Because the harmonic oscillator bases are discrete and finite in truncated BLFQ, the resulting reduced density matrix is likewise discrete and finite. Consequently, the entanglement entropy evaluated in this representation is protected from transverse IR divergences, completely avoiding terms proportional to $\operatorname{log}(L_\perp^2/(2\pi)^2)$. The overall IR parameter, encoded in the 3D box-regularization volume $V$, can therefore be uniquely fixed by matching the entanglement entropy evaluated in the momentum representation to the corresponding finite result obtained in the harmonic oscillator representation.

However, extracting the entropy in the harmonic oscillator representation is computationally demanding. With current computational resources, only a finite range of $\sigma$ values is numerically accessible, and this range strictly excludes the $\sigma\rightarrow0$ regime. A meaningful comparison between the harmonic-oscillator entropy and its momentum-space counterpart therefore requires a rigorous extrapolation to the narrow wave packet limit.

\begin{figure*}
\centering 
\subfigure[\ ]{\includegraphics[width=.32\textwidth]{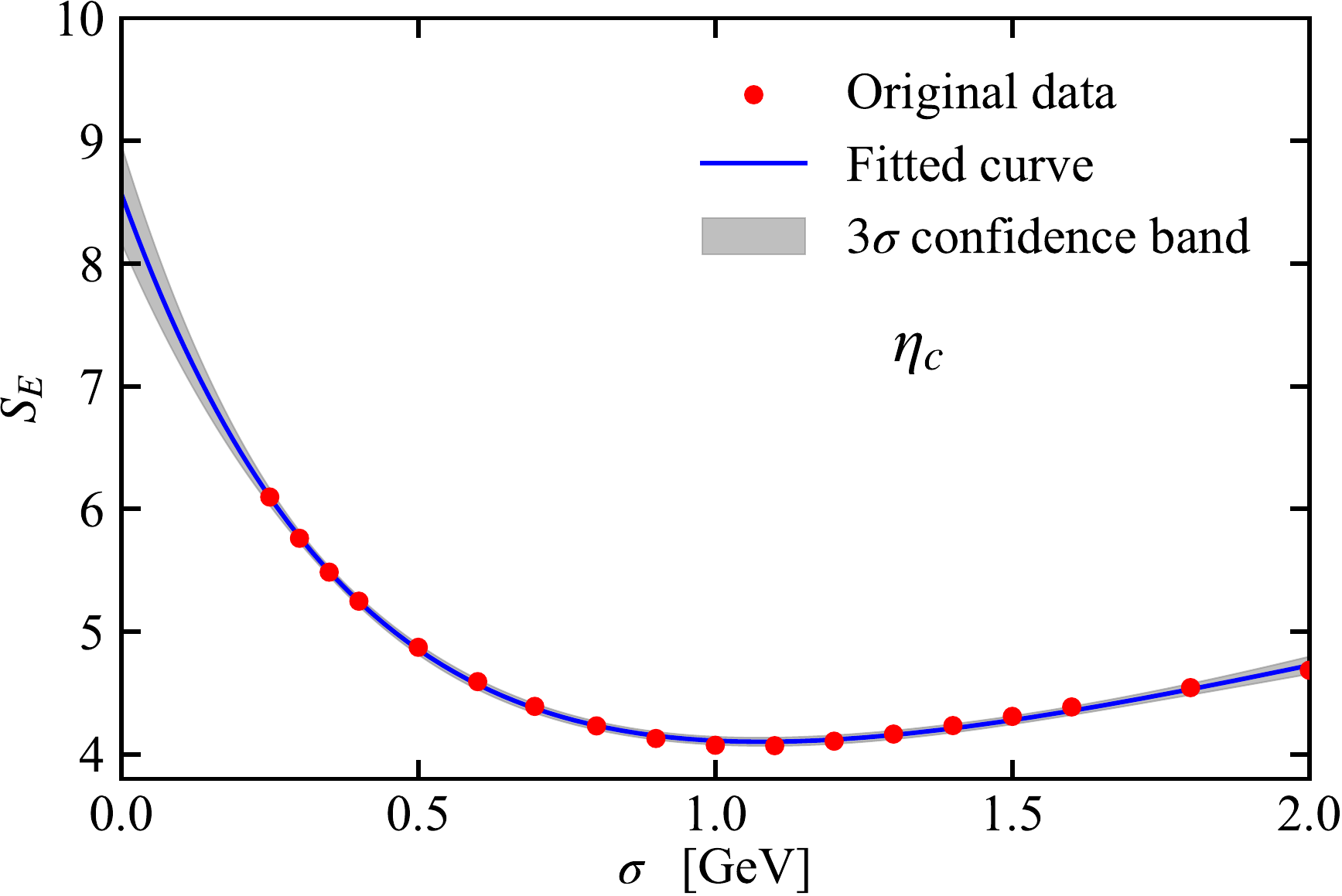}} \hfill
\subfigure[\ ]{\includegraphics[width=.32\textwidth]{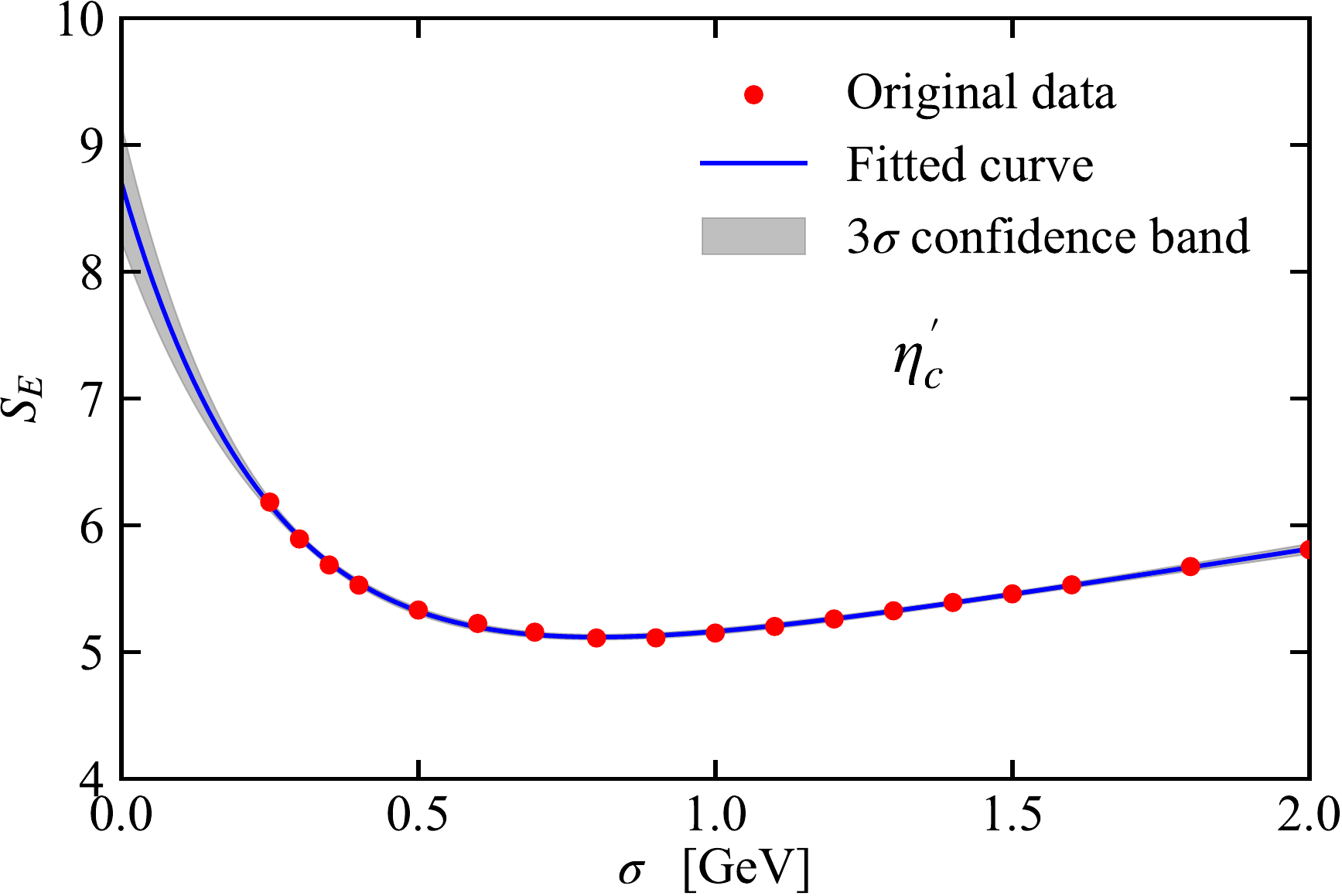}} \hfill
\subfigure[\ ]{\includegraphics[width=.32\textwidth]{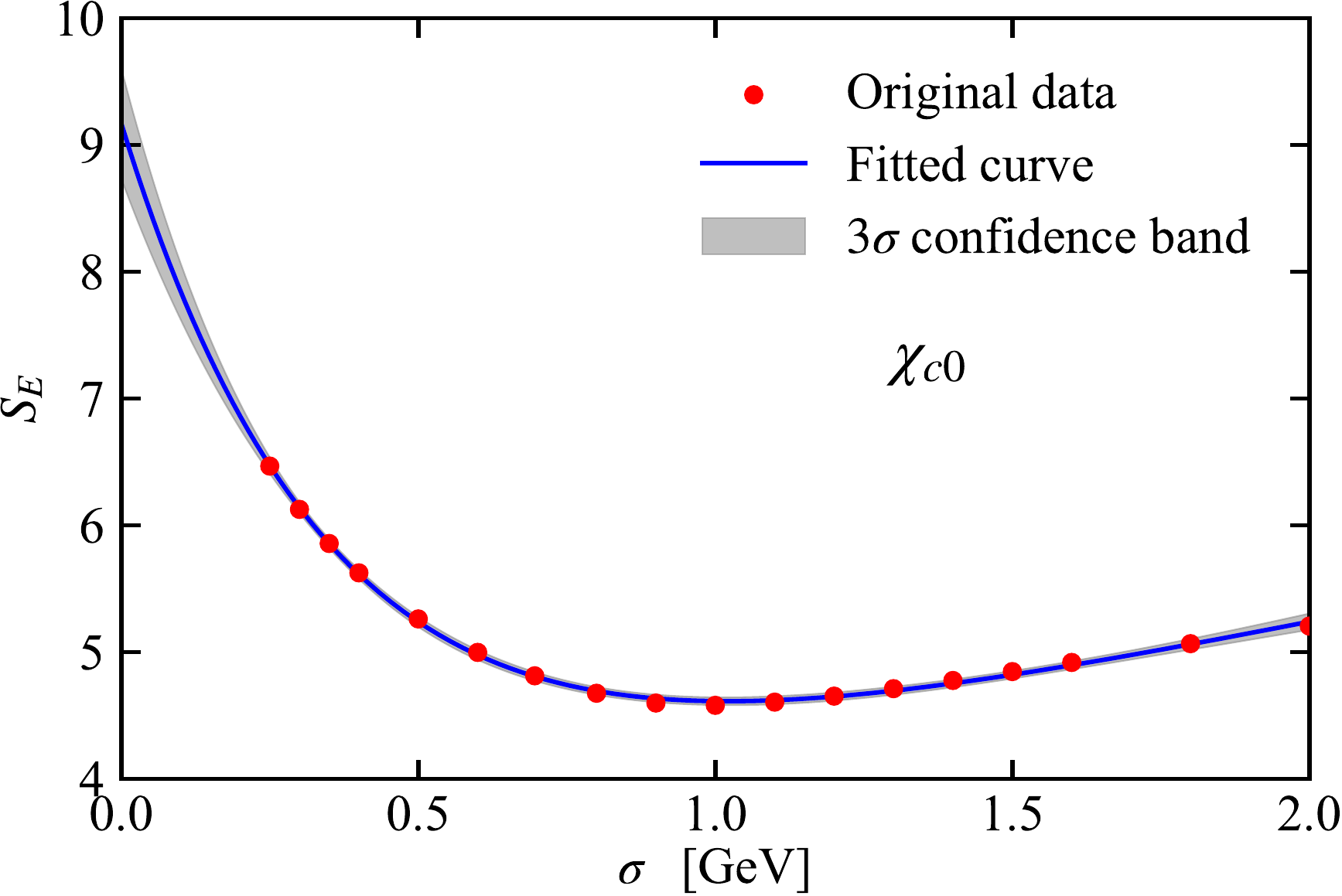}}
\caption{Entanglement entropy between quark and anti-quark d.o.f's in charmonium as a function of Gaussian wave packet width $\sigma$ in the transverse direction. The entanglement entropy is calculated in the harmonic oscillator representation with $N_\text{max}=8$. The red scatter points represent the calculated results, the blue solid lines are the nonlinear fitting results, the gray bands denote the $99.73\%$ confidence interval. The nonlinear fitting function is given by $f(x)=ae^{-bx}+cx+d$.}
\label{fig:HO Entropy}
\end{figure*}

\begin{table*}[t]
\centering
\caption{Nonlinear fitting results and extrapolation outcomes at $\sigma=0$.}
\begin{tabular}{cccc}
\toprule
& extrapolation at $\sigma=0$ & $99.73\%$ confidence band & adjusted $R^2$\\
\midrule
$\eta_c$ & 8.5711 & $[8.1712, 8.9710]$ & 0.9983 \\
$\eta_c'$ & 8.7059 & $[8.2424, 9.1693]$ & 0.9973 \\
$\chi_{c0}$ & 9.1702 & $[8.7351, 9.6052]$ & 0.9982 \\
\bottomrule
\end{tabular}
\label{tab:HO Entropy}
\end{table*}

\begin{table}[t]
\centering
\caption{Entanglement entropy for charmonium obtained from Eq.~(\ref{quark entropy}) without the IR term $\operatorname{log}(P_0^+V/(2\pi)^3)$.}
\begin{tabular}{cccc}
\toprule
& $\eta_c$ & $\eta'_c$ & $\chi_{c0}$ \\
\midrule
$S_E$ & $2.73743$ & $2.70070$ & $3.10068$ \\
\bottomrule
\end{tabular}
\label{tab:charmonium entropy}
\end{table}

Figure~\ref{fig:HO Entropy} presents the quark-antiquark entanglement entropy in charmonium, evaluated within the harmonic oscillator representation. These calculations are performed across a range of transverse Gaussian wave packet widths, $\sigma$, at a fixed harmonic oscillator basis truncation of $N_{\text{max}}=8$. For the longitudinal c.m. motion defined in Eq.~(\ref{eqn:center-of-mass motion}), we impose the narrow wave packet limit by setting 
\begin{equation}
    P_0^+L = 2N_\text{max}\times2\pi = 32\pi,
\end{equation}
while retaining a finite width $\sigma$ in the transverse direction.

Table~\ref{tab:HO Entropy} summarizes the nonlinear fitting results and the corresponding extrapolations to the strict narrow wave packet limit ($\sigma\rightarrow 0$). By matching these extrapolated values to the momentum-space entanglement entropy [Eq.~(\ref{quark entropy})] with the IR divergent term subtracted (see Table~\ref{tab:charmonium entropy}), we uniquely determine the IR cutoff and the transverse box-regularization parameter:
\begin{equation}
    \Lambda_{\text{IR}} \doteq \frac{\kappa}{\sqrt{3N_{\text{max}}}}, \qquad
    L_{\perp} = \frac{2\pi}{\Lambda_{\text{IR}}} \doteq \frac{2\pi\sqrt{3N_{\text{max}}}}{\kappa}.
\label{eqn:IR parameter}
\end{equation}
Furthermore, the functional behavior shown in Figure~\ref{fig:HO Entropy} reveals a non-monotonic dependence on the transverse wave packet width: as $\sigma$ increases, the entanglement entropy initially drops rapidly before undergoing a gradual rise. The specific value of $\sigma$ at which the entropy reaches its minimum varies slightly depending on the hadronic state, but consistently falls near $\sigma \approx 1\,\mathrm{GeV}$. 

\subsection{Entanglement entropy}

\begin{figure}
\centering 
\subfigure[\ ]{\;\;\includegraphics[width=.39\textwidth]{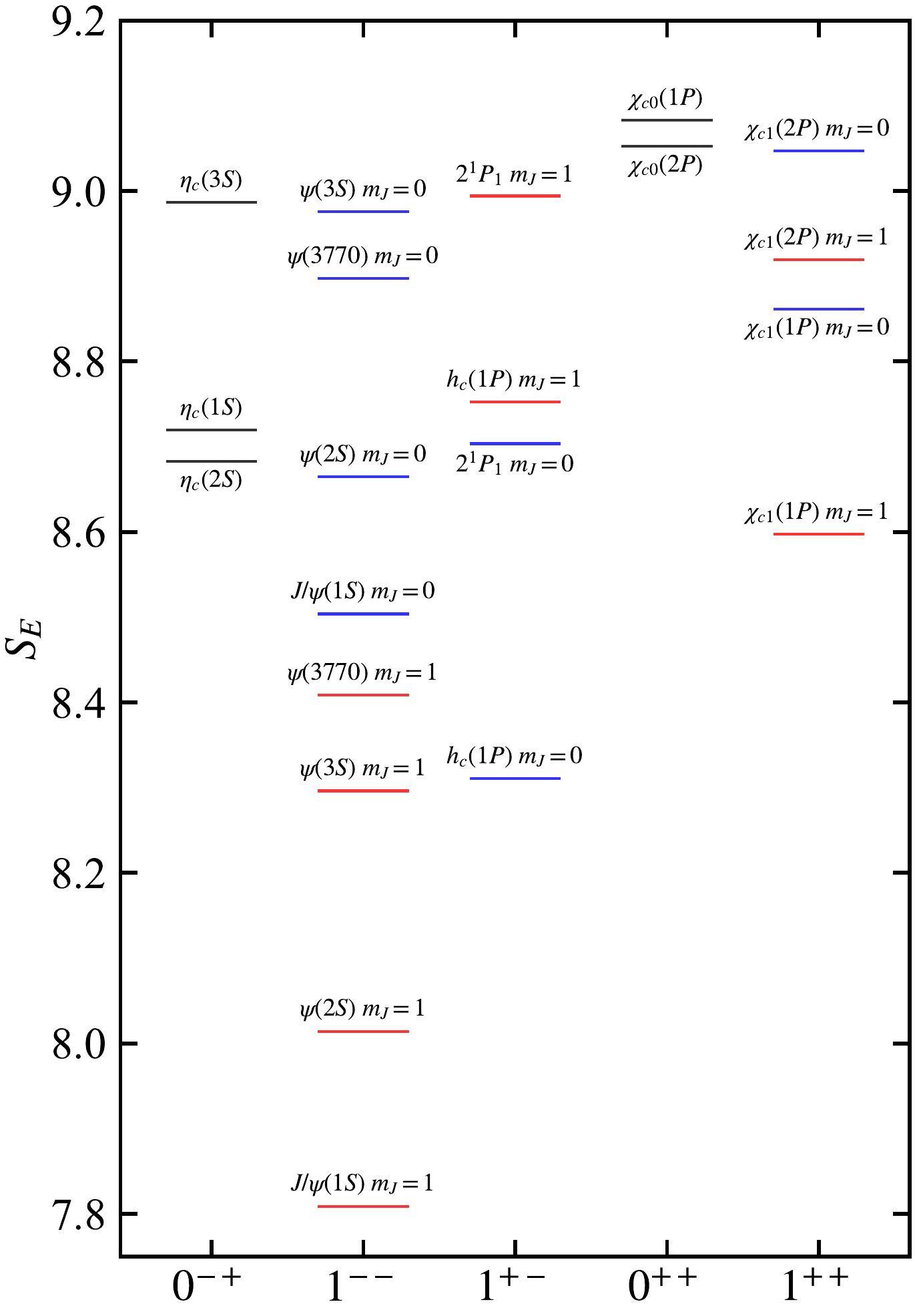}} \\
\subfigure[\ ]{\includegraphics[width=.4\textwidth]{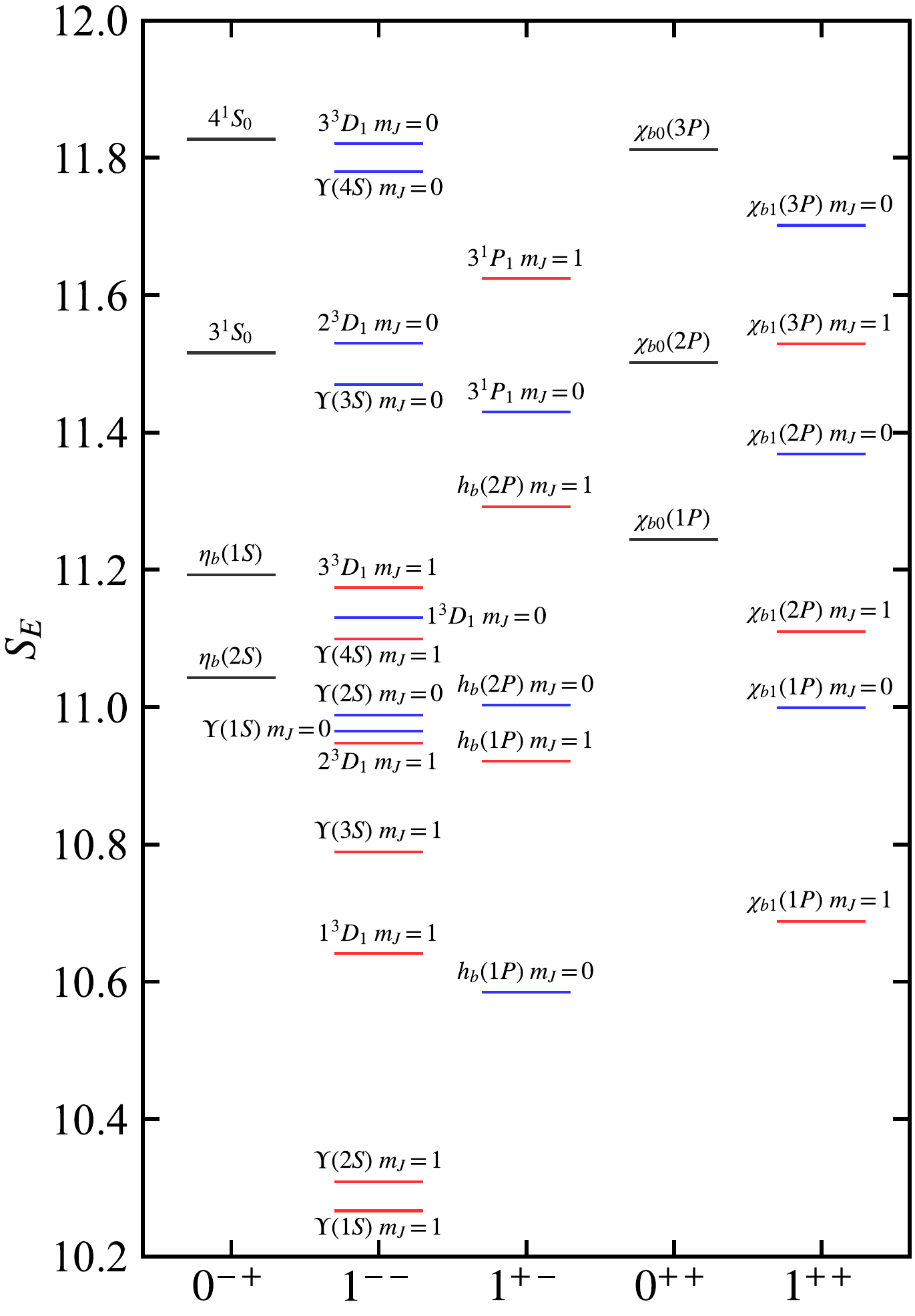}}
\caption{The entanglement entropy between quark and anti-quark d.o.f's for charmonium (\textit{top panel}) with $P_0^+V/(2\pi)^3=396\ [\text{GeV}^{-2}]$ and bottomonium (\textit{bottom panel}) with $P_0^+V/(2\pi)^3=3186\ [\text{GeV}^{-2}]$.}
\label{fig:Entropy spectrum}
\end{figure}

We evaluate the entanglement entropy between quark and antiquark d.o.f.'s for charmonium and bottomonium using the LFWFs obtained within the BLFQ framework. In particular, for spin-1 quarkonia, we present the entanglement entropy for the eigenstates $\ket{m_J=0}$ and $\ket{m_J=1}$. The corresponding entropy spectrum is illustrated in Figure~\ref{fig:Entropy spectrum}. The IR parameters are fixed according to Eq.~(\ref{eqn:IR parameter}). For charmonoium, we adopt the truncation $N_\text{max}=8$, which yields $P_0^+V/(2\pi)^3=396\ [\text{GeV}^{-2}]$. For bottomonium, we adopt the truncation $N_\text{max}=32$, which yields $P_0^+V/(2\pi)^3=3186\ [\text{GeV}^{-2}]$. Our results show that the entanglement entropy of quarkonium does not exhibit a clear systematic correlation with the radial or orbital excitation of the state. For spin-1 quarkonia, however, the entanglement entropy differs significantly between $\ket{m_J=0}$ and $\ket{m_J=1}$, indicating a strong dependence of partonic entanglement on the polarization state of the meson. Specifically, for the $1^{--}$ and $1^{++}$ quarkonia, the $m_J=0$ state has a larger entanglement entropy than the $m_J=1$ state, whereas for the $1^{+-}$ quarkonia, the opposite trend is observed.

\begin{figure}
\centering 
\subfigure[\ ]{\includegraphics[width=.23\textwidth]{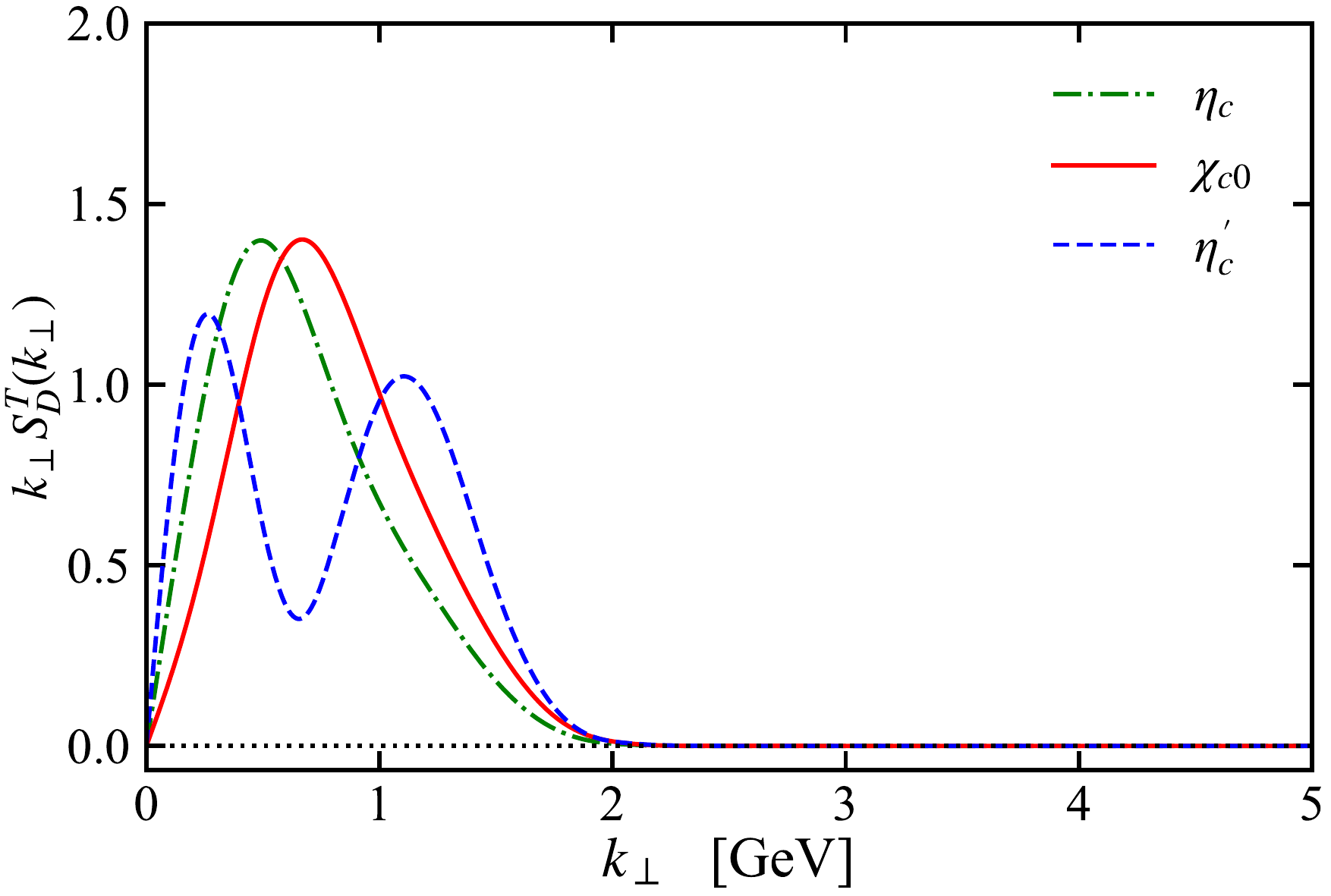}} \hfill
\subfigure[\ ]{\includegraphics[width=.23\textwidth]{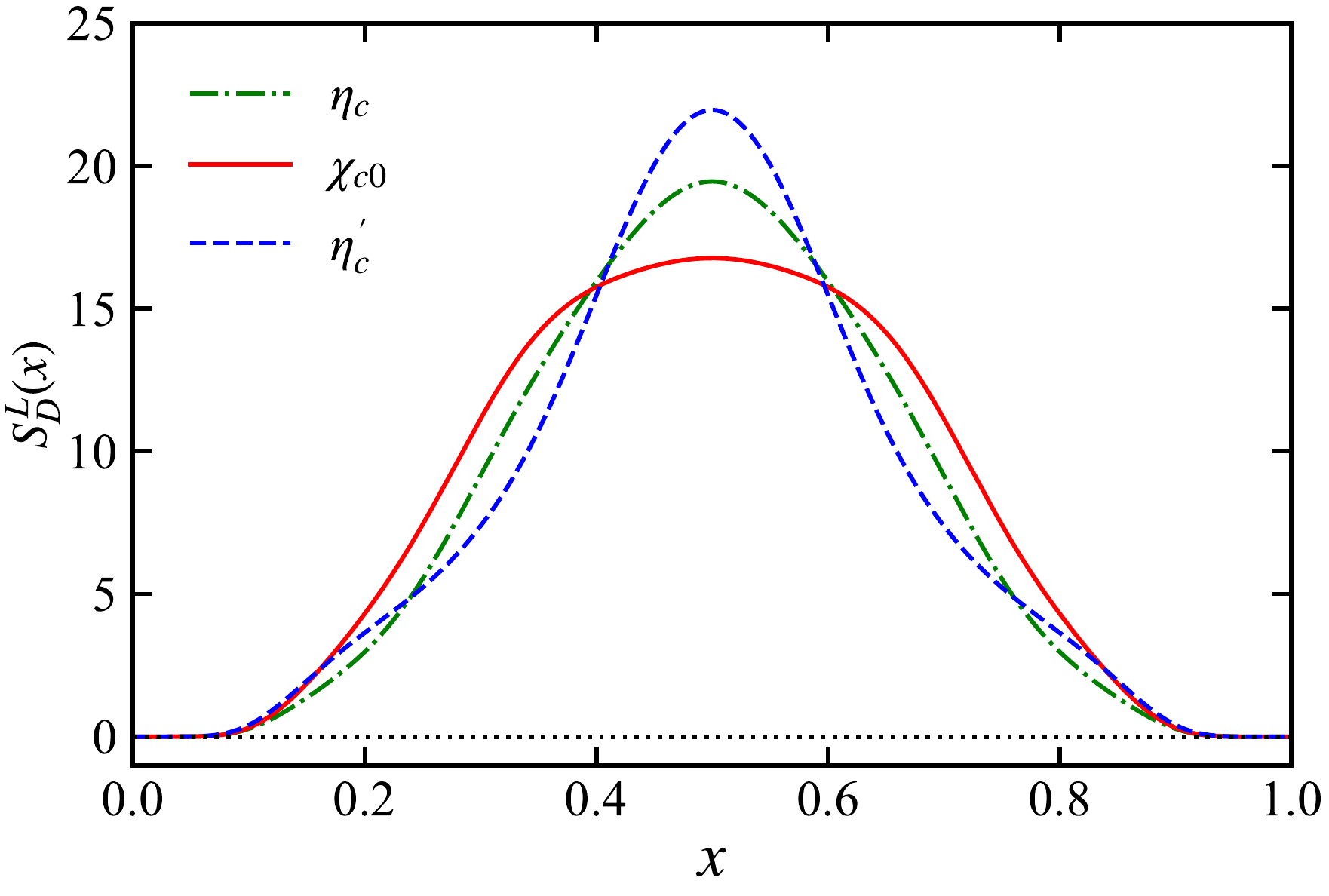}} \hfill
\subfigure[\ ]{\includegraphics[width=.23\textwidth]{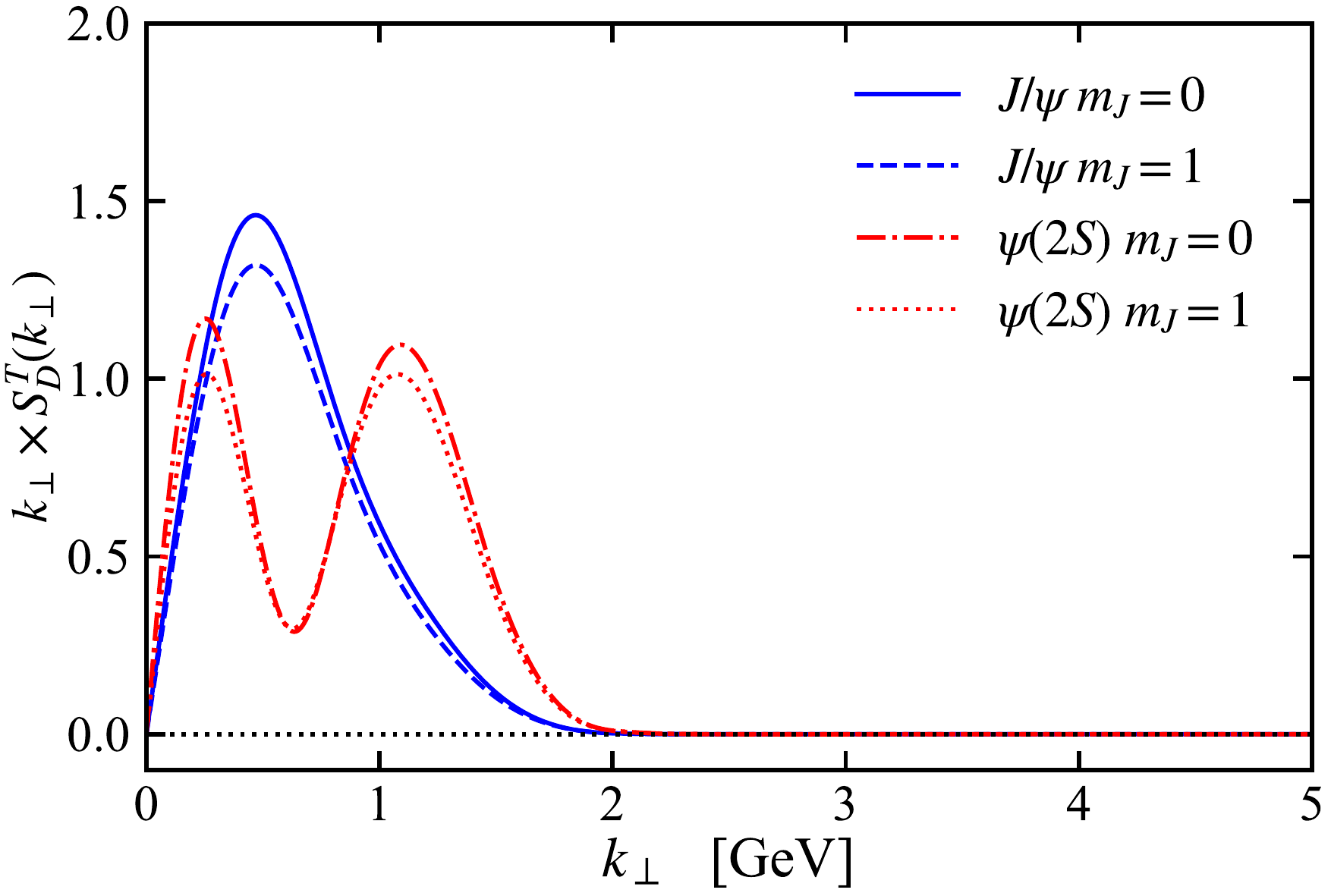}} \hfill
\subfigure[\ ]{\includegraphics[width=.23\textwidth]{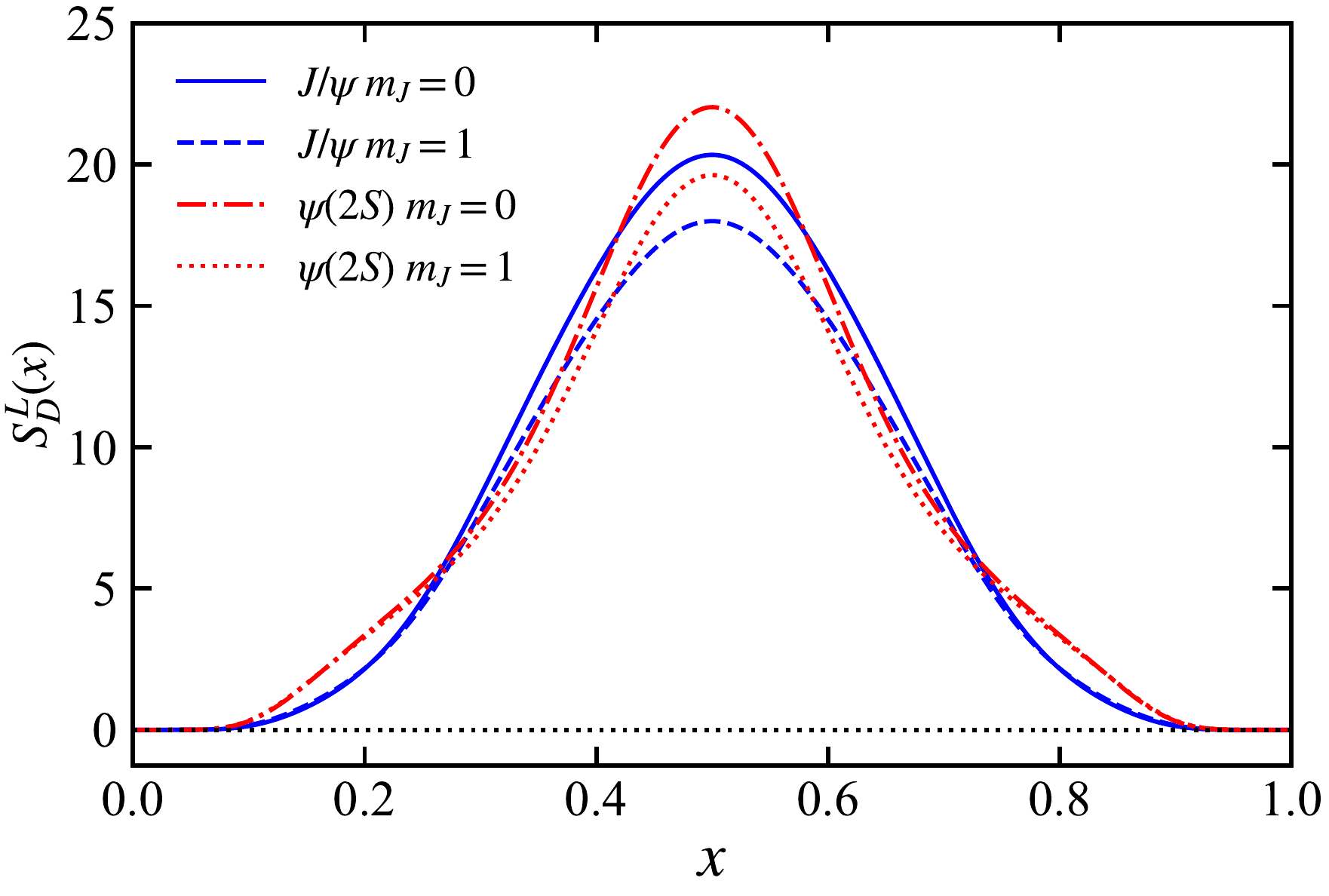}} \hfill
\subfigure[\ ]{\includegraphics[width=.23\textwidth]{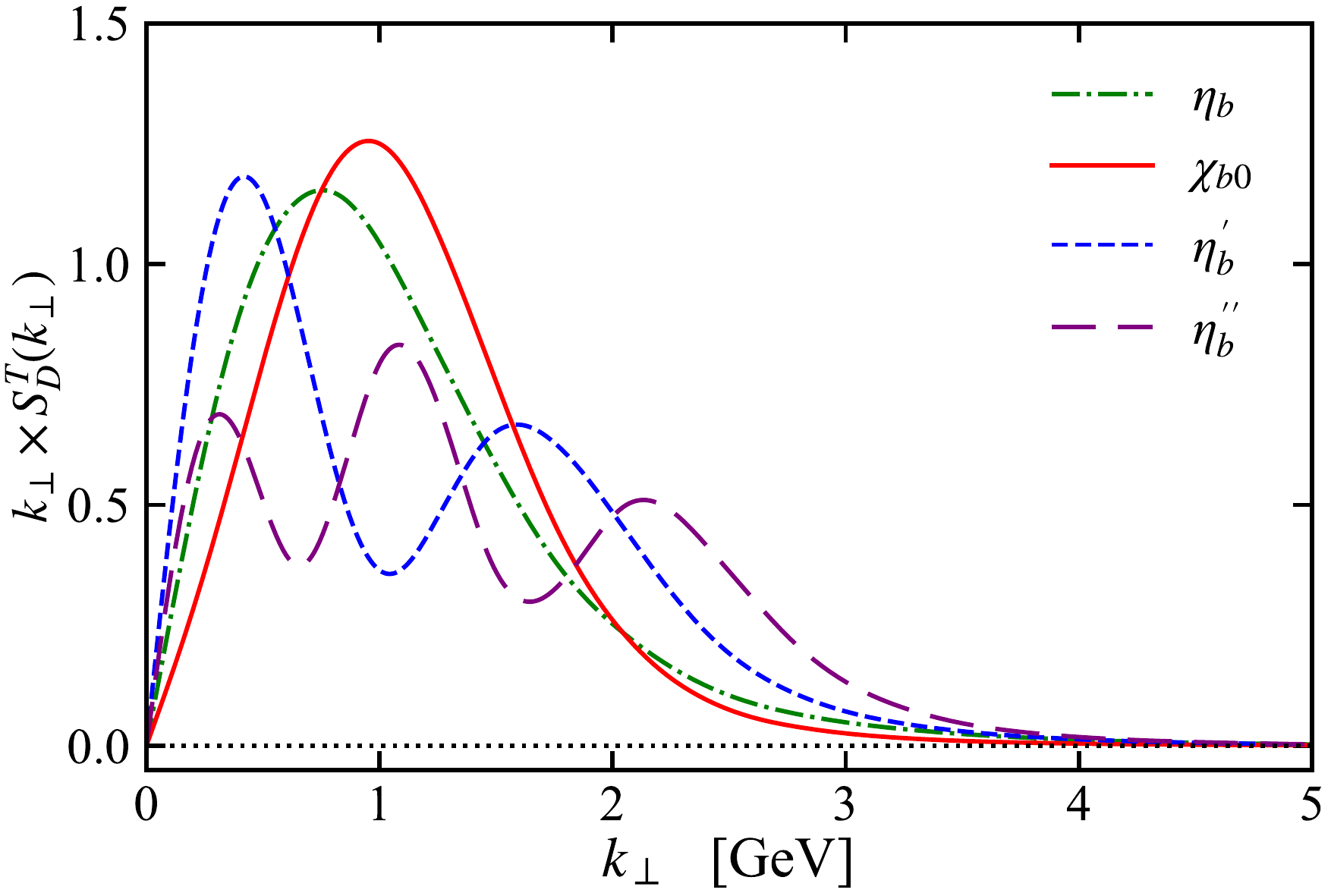}} \hfill
\subfigure[\ ]{\includegraphics[width=.23\textwidth]{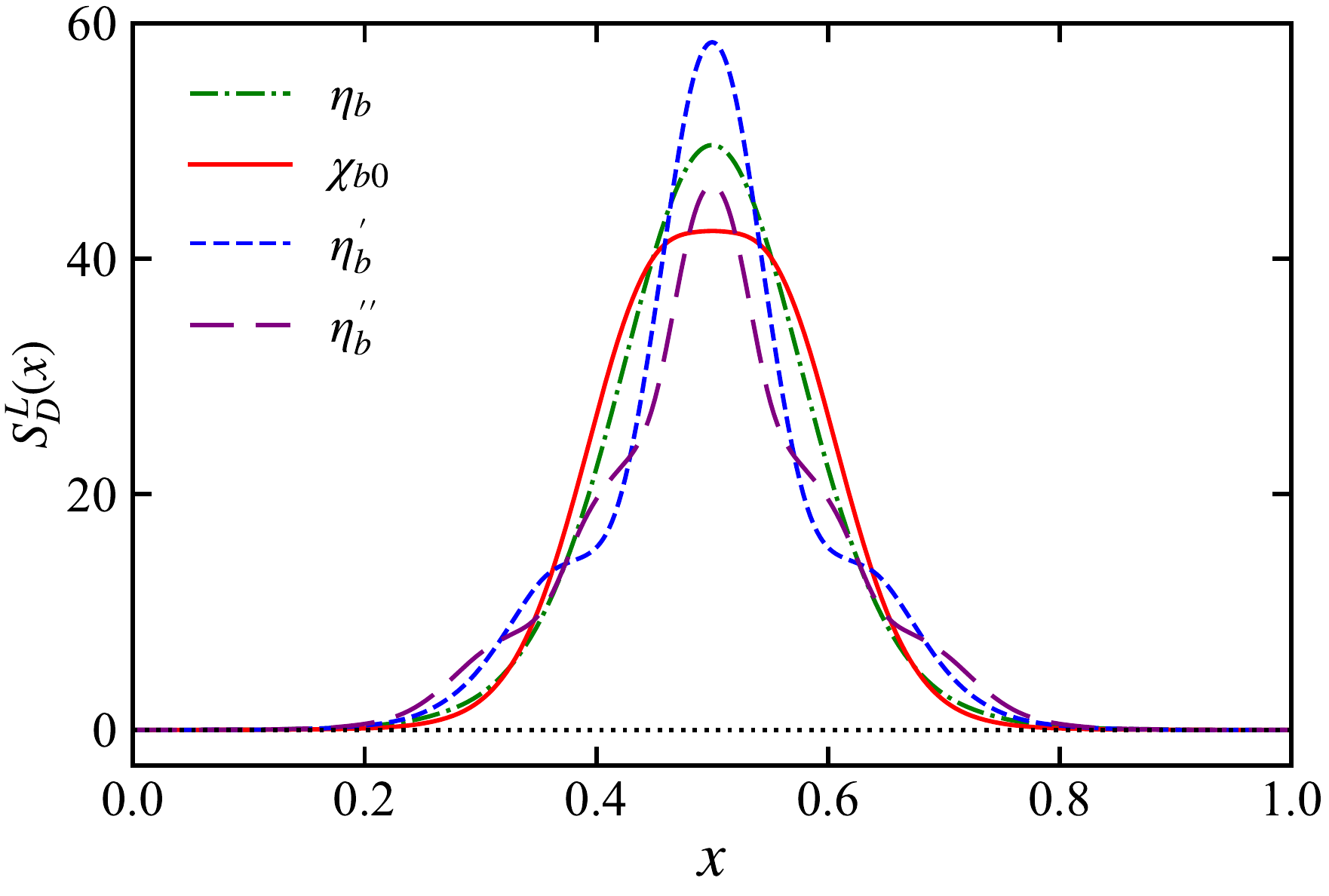}} \hfill
\subfigure[\ ]{\includegraphics[width=.23\textwidth]{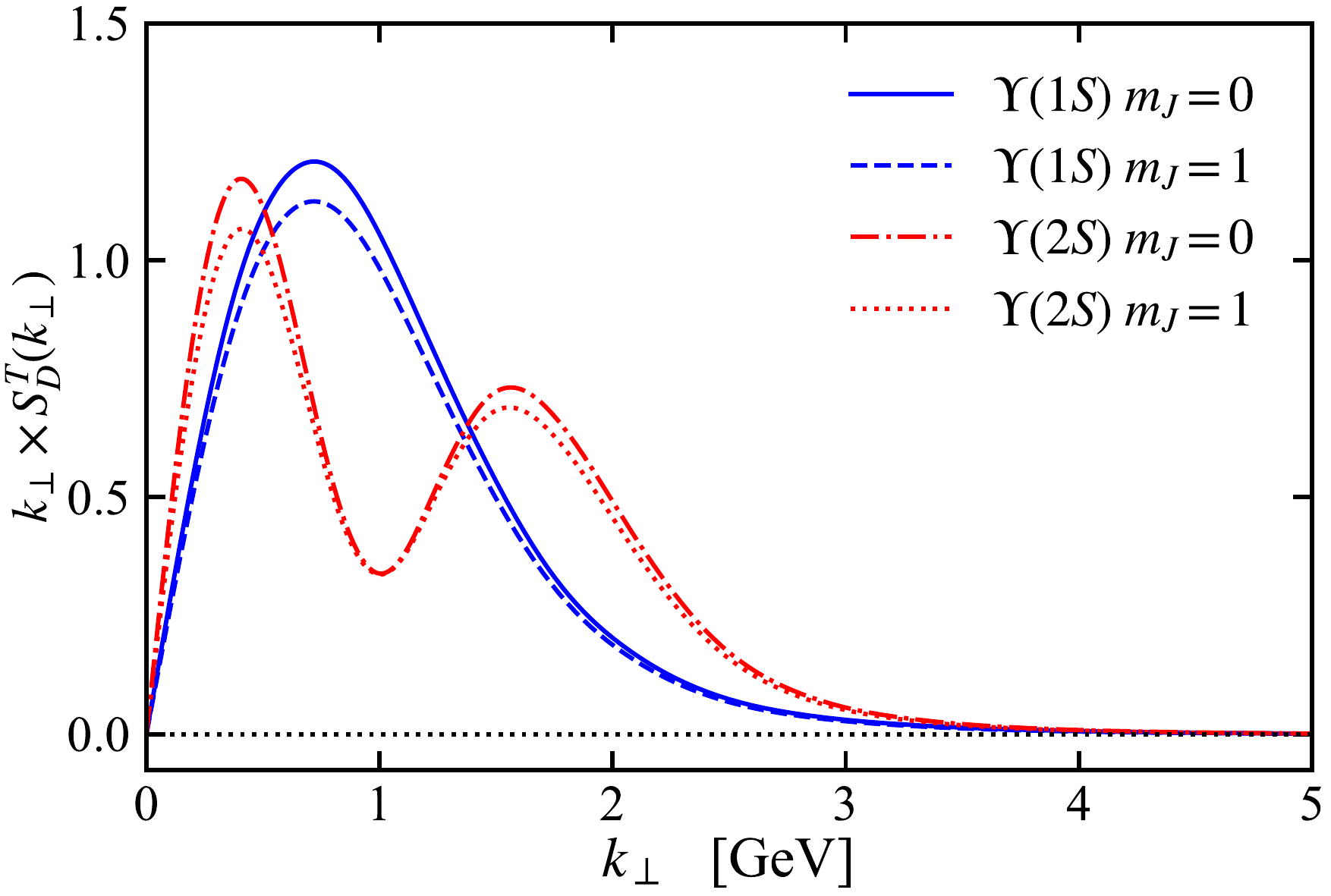}} \hfill
\subfigure[\ ]{\includegraphics[width=.23\textwidth]{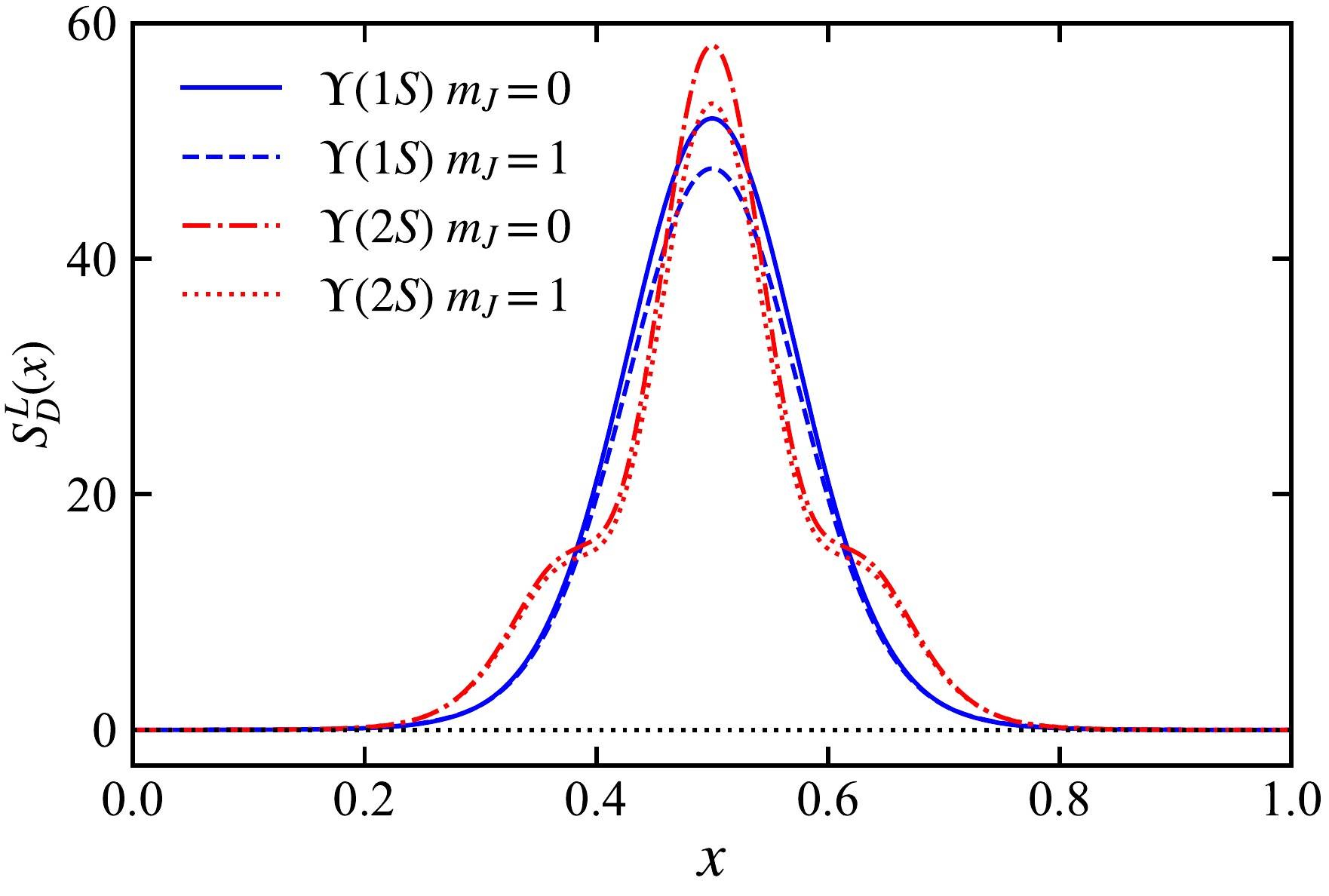}}
\caption{Distribution of entanglement entropy in transverse and longitudinal momentum for charmonium (panels (a)-(d)) with $P_0^+V=396\ \text{GeV}^{-2}$ and bottomonium (panels (e)-(h)) with $P_0^+V=3186\ \text{GeV}^{-2}$.}
\label{fig:Entropy density}
\end{figure}

To further characterize the momentum-space distribution of entanglement, we introduce the transverse entropy density
\begin{multline}
    S_D^T(\vec{k}_{\perp};\Lambda)=-\sum_{a=1}^2\int\text{d}x\\ 
    \times \lambda_a^{\Lambda}(x,\vec{k}_{\perp})\ \text{log}\Bigg[\frac{(2\pi)^3}{P_0^+V}\lambda_a^{\Lambda}(x,\vec{k}_{\perp})\Bigg],
\end{multline}
and the longitudinal entropy density
\begin{multline}
    S_D^L(x;\Lambda)=-\sum_{a=1}^2\int\text{d}^2 {k}_{\perp}\\ 
    \times \lambda_a^{\Lambda}(x,\vec{k}_{\perp})\ \text{log}\Bigg[\frac{(2\pi)^3}{P_0^+V}\lambda_a^{\Lambda}(x,\vec{k}_{\perp})\Bigg].
\end{multline}
For spin-0 quarkonia, one has $\lambda_{1,2}(x,k_\perp)=f_1(x,k_\perp)$.
The entropy densities defined above do not include the contribution from color d.o.f.'s, since color entanglement is a global contribution shared by the quark-antiquark pair and cannot be decomposed into a momentum-space entropy density.
Figure~\ref{fig:Entropy density} shows the entropy densities for spin-0 and spin-1 quarkonia. For the transverse entropy density, the number of peaks is closely related to the excitation pattern of the state. By contrast, the longitudinal entropy density peaks at $x=0.5$ as a consequence of the symmetry between the quark and antiquark, which also mirrors the evolution of the PDF shape~\cite{Li:2017mlw}. Moreover, for spin-1 quarkonia, although the total entanglement entropy differs significantly between the $m_J=0$ and $m_J=1$ states, their transverse and longitudinal entropy densities display qualitatively similar trends.

\begin{figure}
\centering
\subfigure[\ ]{\includegraphics[width=.23\textwidth]{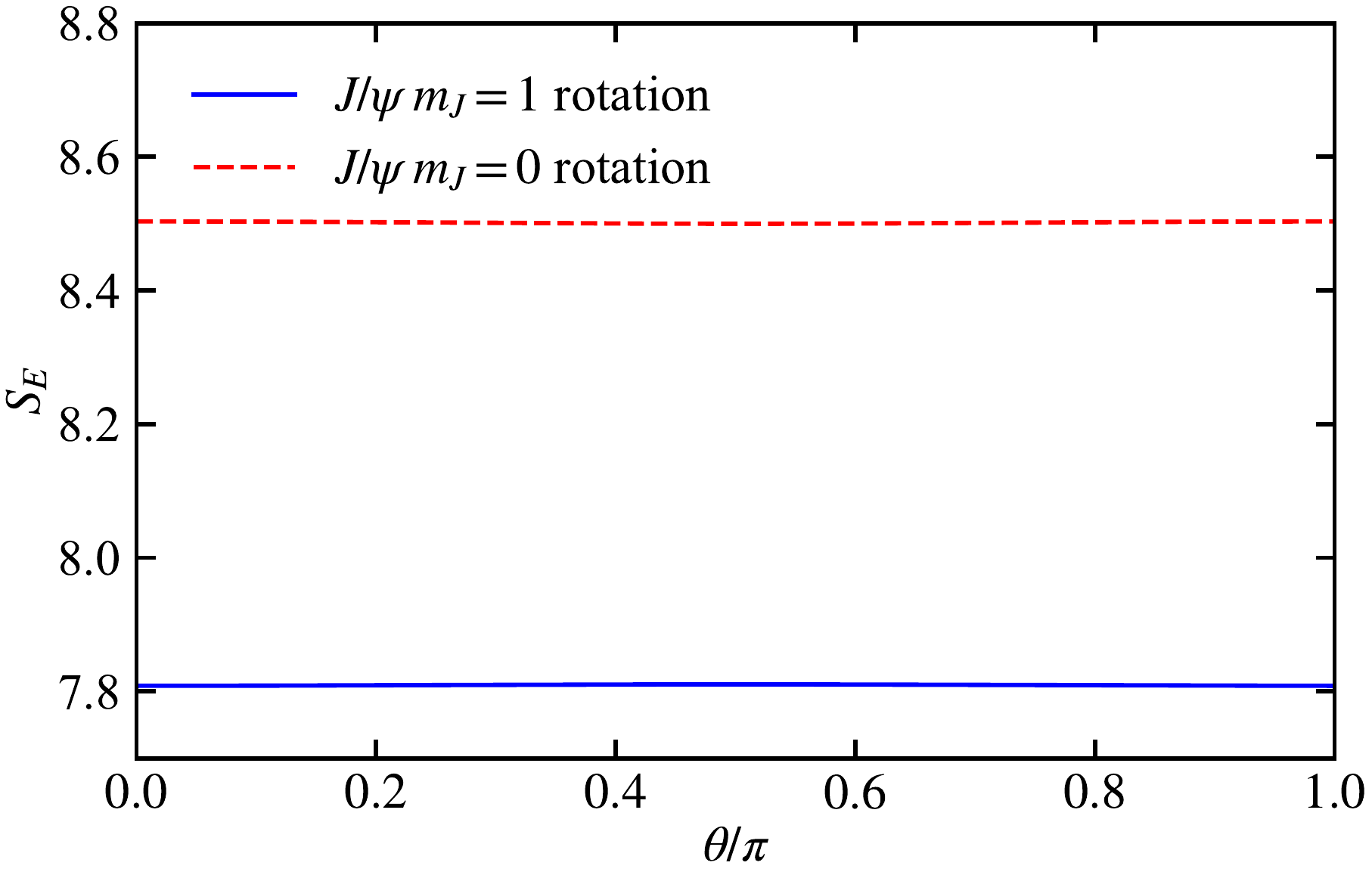}} \hfill
\subfigure[\ ]{\includegraphics[width=.23\textwidth]{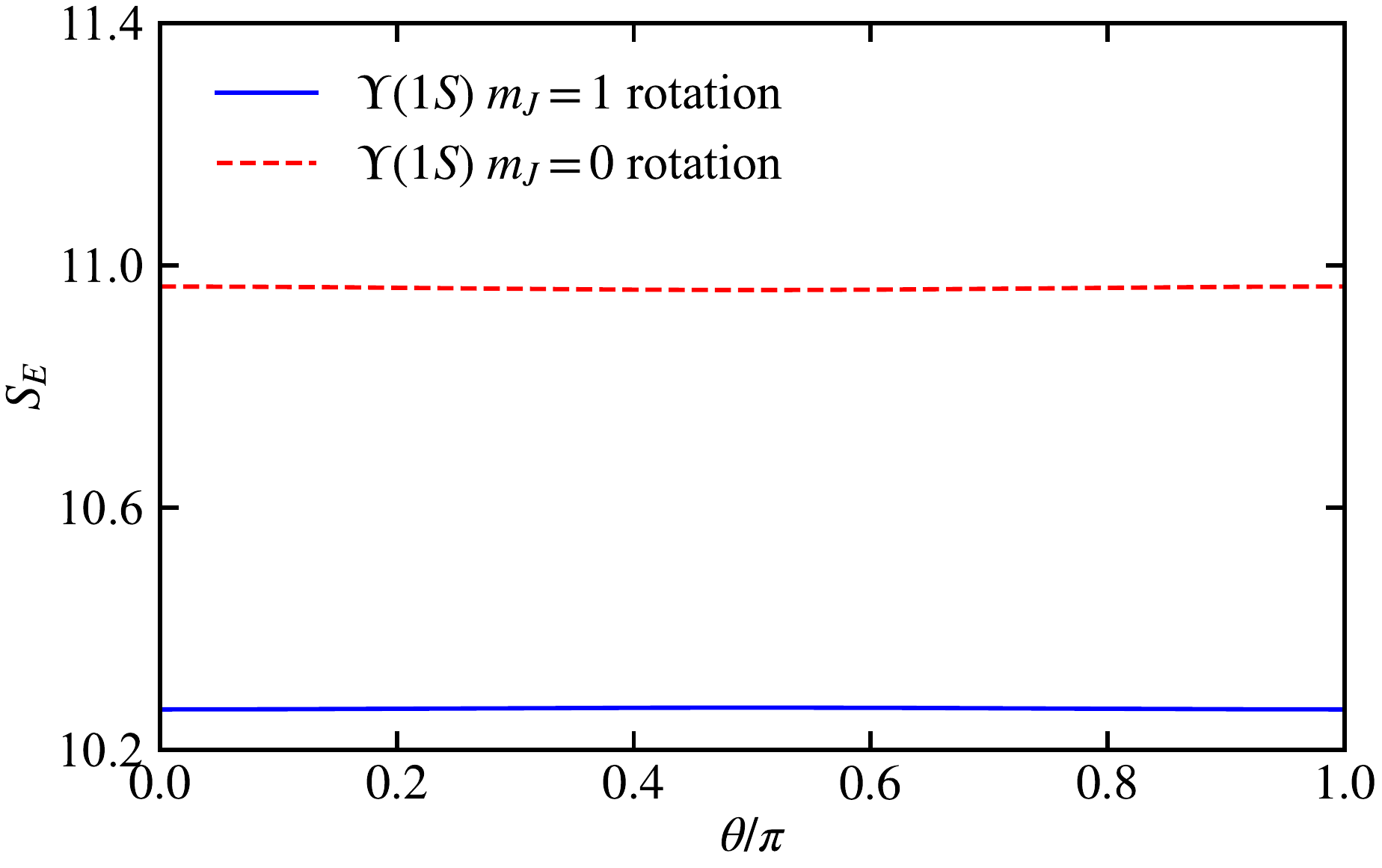}} \hfill
\subfigure[\ ]{\includegraphics[width=.23\textwidth]{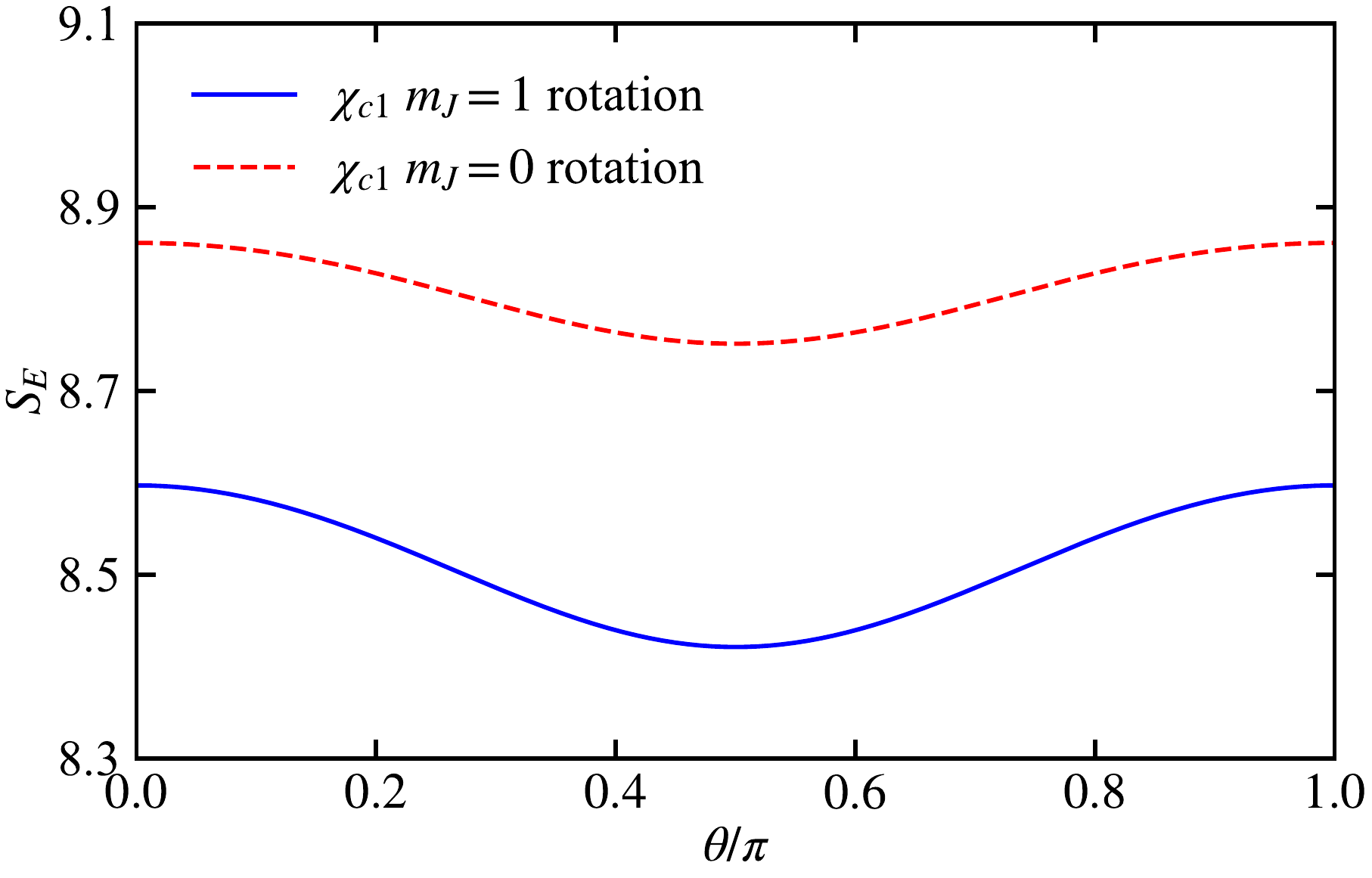}} \hfill
\subfigure[\ ]{\includegraphics[width=.23\textwidth]{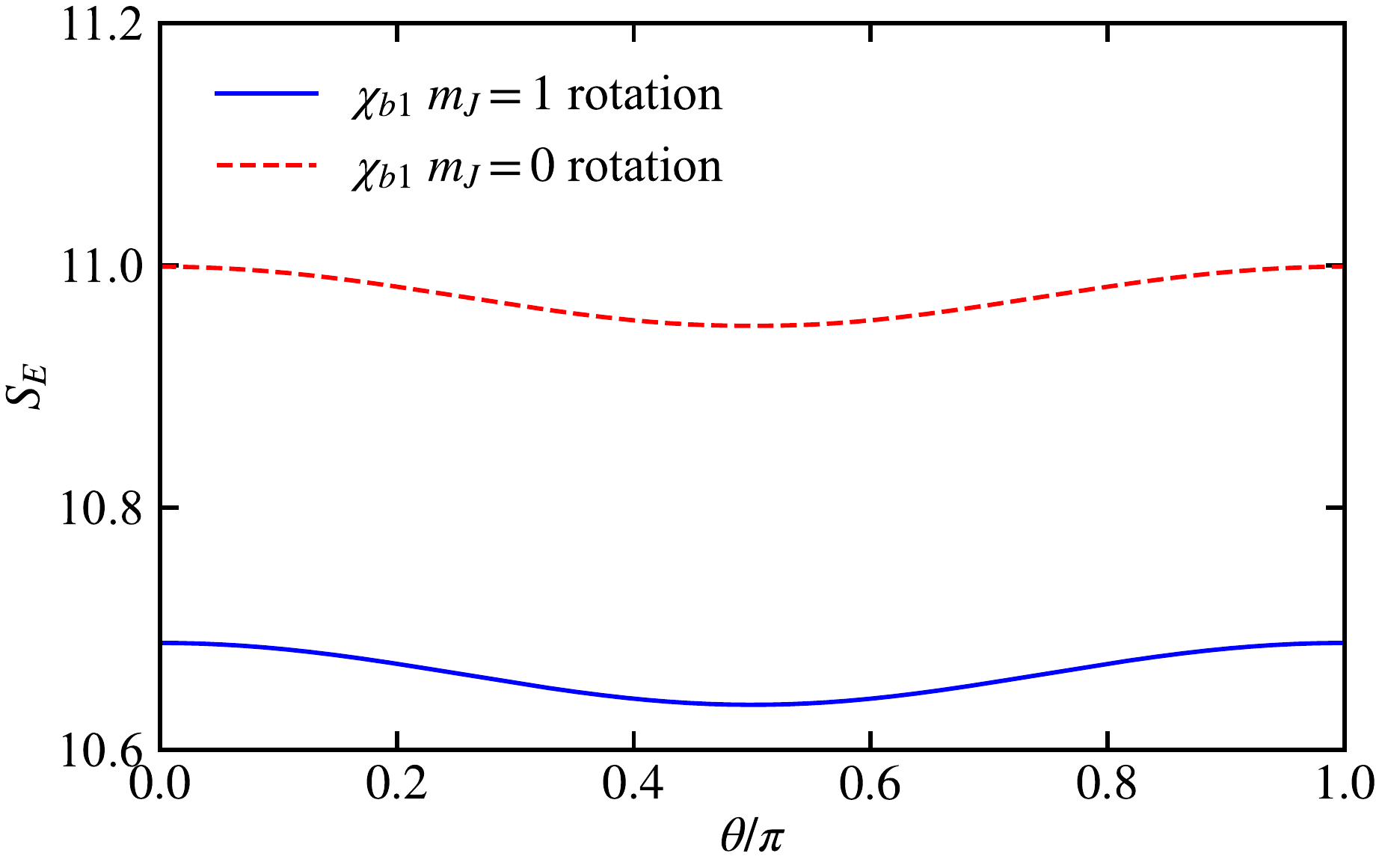}} \hfill
\subfigure[\ ]{\includegraphics[width=.23\textwidth]{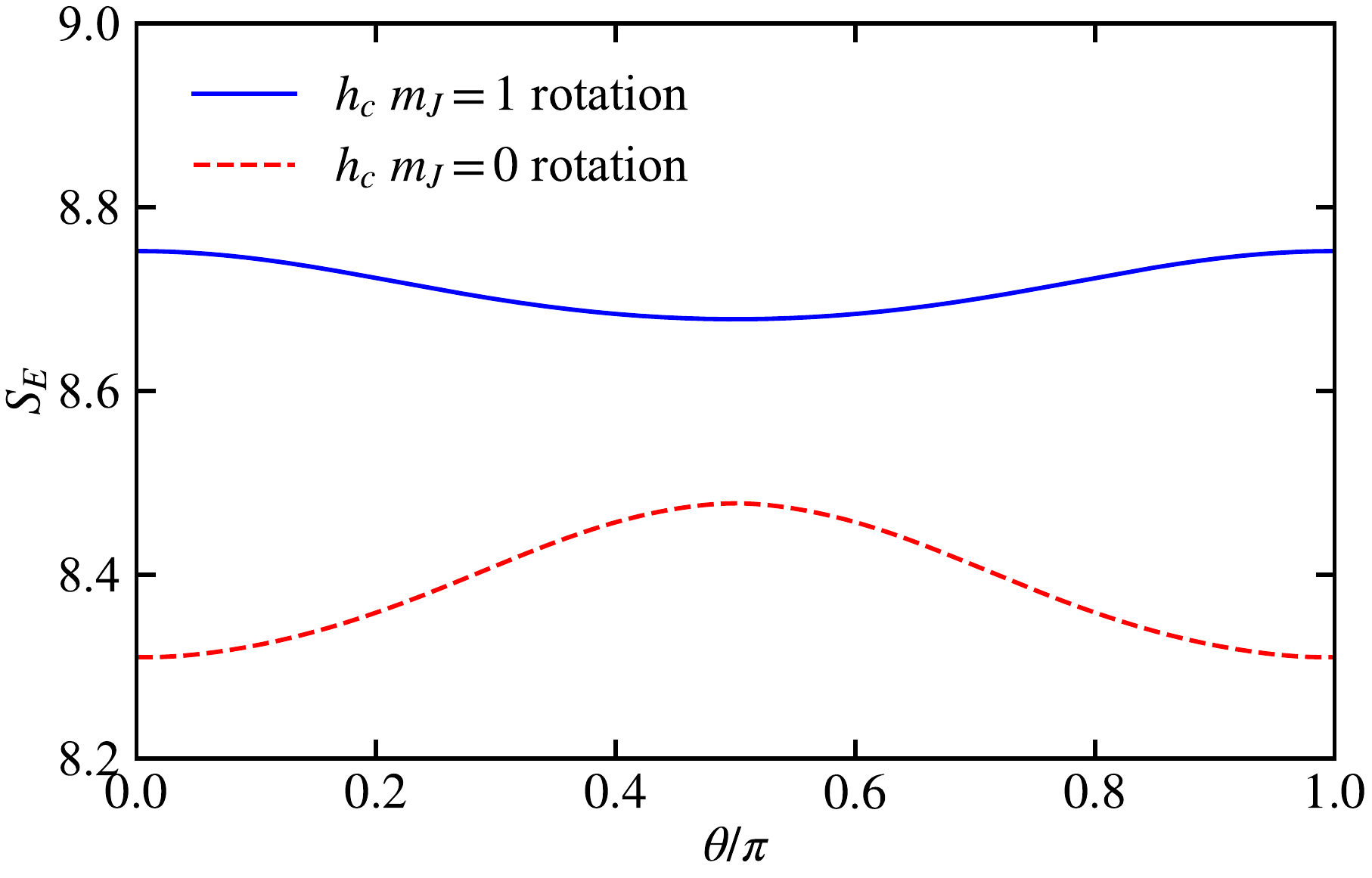}} \hfill
\subfigure[\ ]{\includegraphics[width=.23\textwidth]{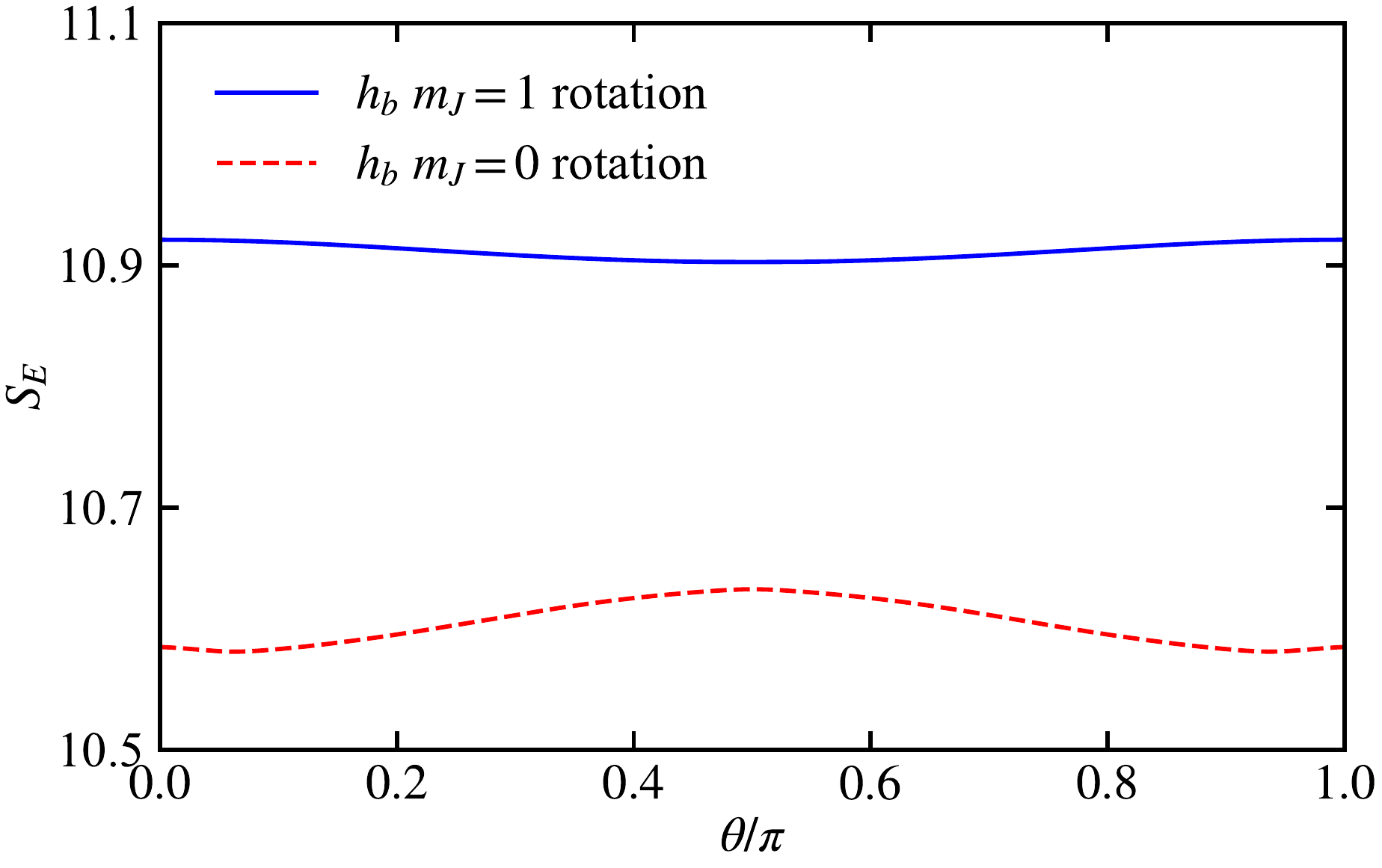}} \hfill
\subfigure[\ ]{\includegraphics[width=.23\textwidth]{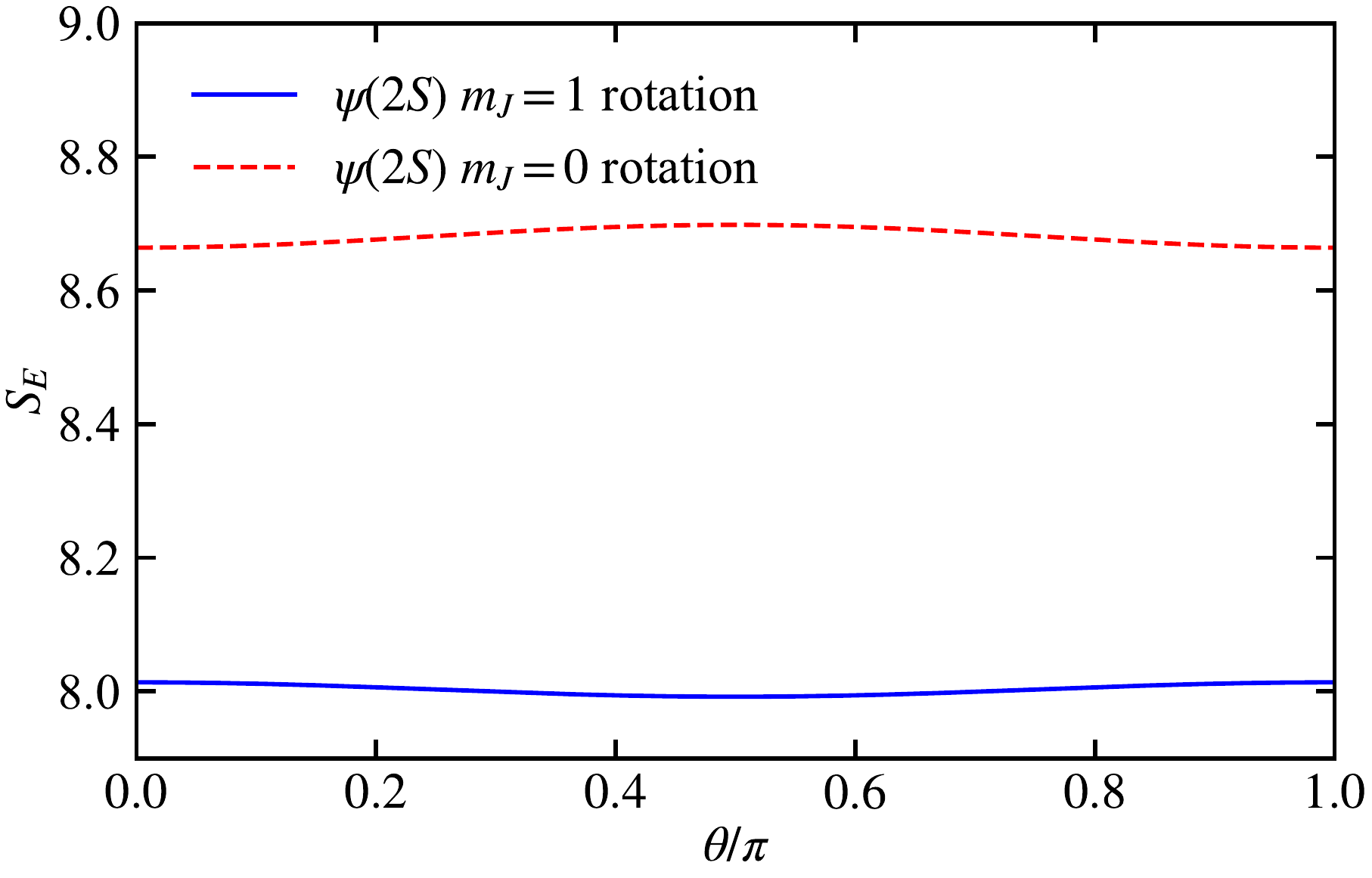}} \hfill
\subfigure[\ ]{\includegraphics[width=.23\textwidth]{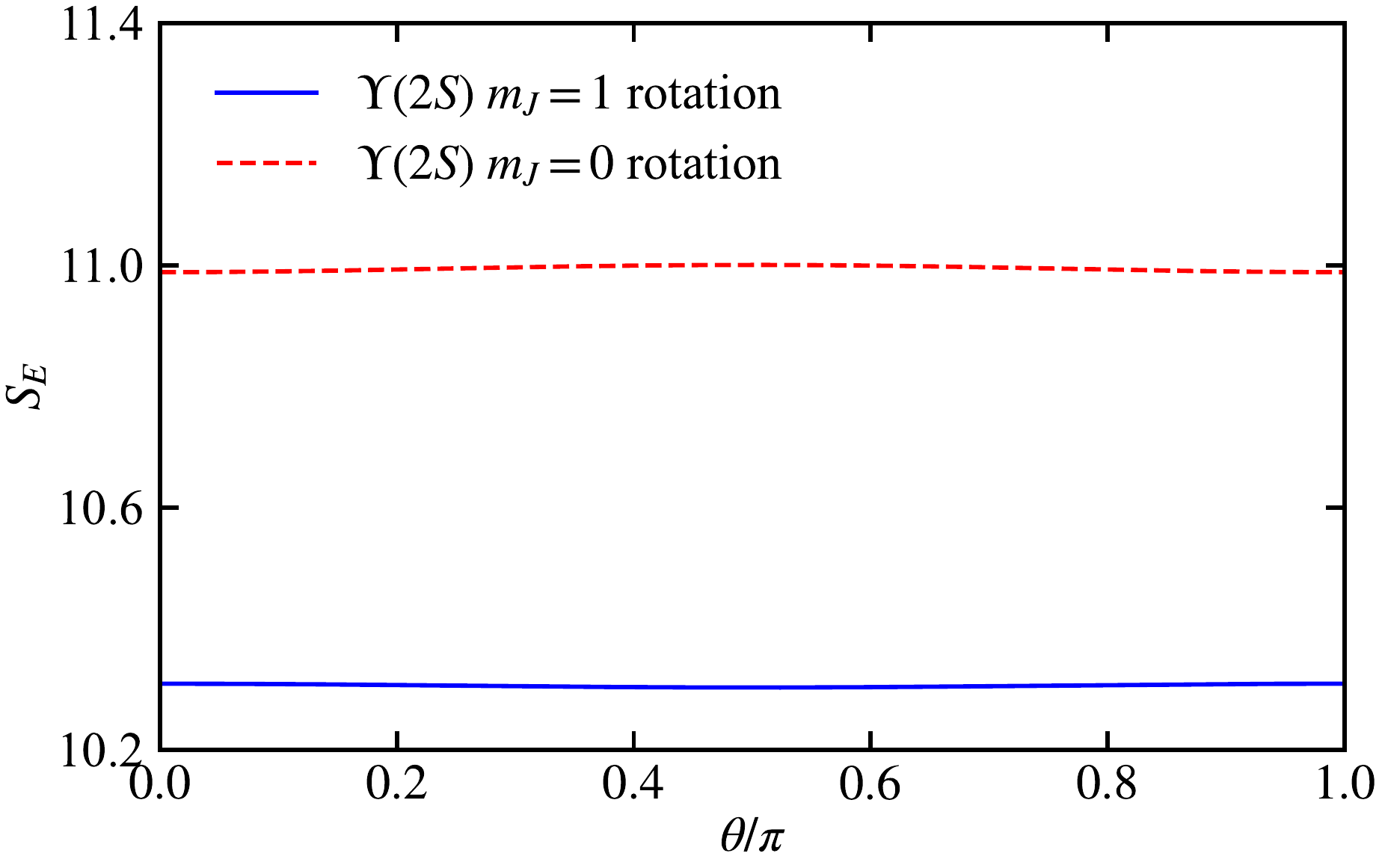}} \hfill
\subfigure[\ ]{\includegraphics[width=.23\textwidth]{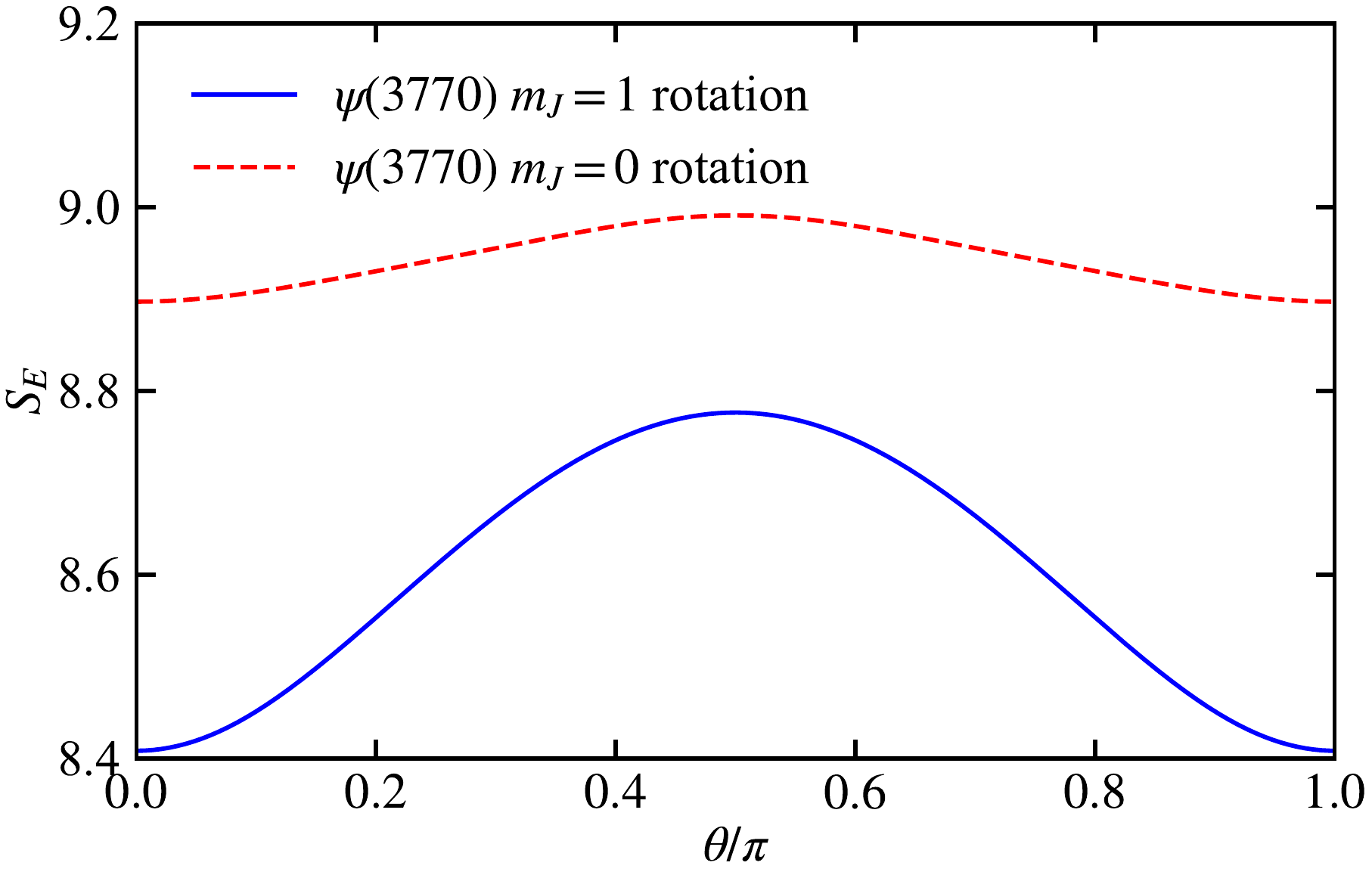}} \hfill
\subfigure[\ ]{\includegraphics[width=.23\textwidth]{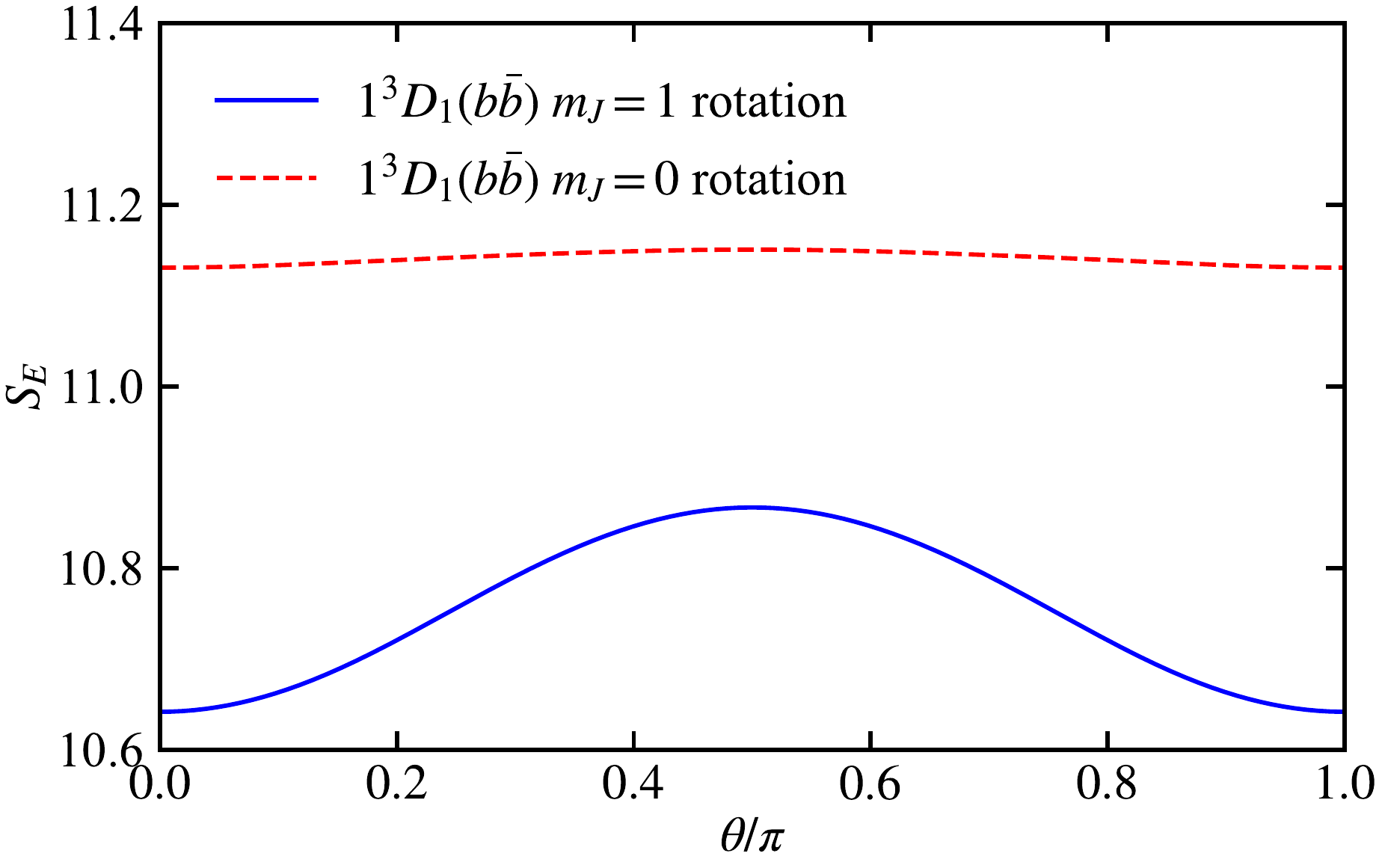}} \hfill
\caption{The entanglement entropy as a function of $\theta$ for spin-1 quarkonium, where $\theta$ is the rotation angle of the $\ket{m_J=0,1}$ state about the $\hat{y}$-axis.}
\label{fig:Angle dependence entropy}
\end{figure}

To investigate the polarization dependence of the entanglement entropy for spin-1 quarkonia, we further compute the entropy for different rotation angles by rotating the $\ket{m_J=0}$ and $\ket{m_J=1}$ states about an axis in the transverse plane. Owing to the symmetry of the transverse plane in light-front dynamics, rotations about different transverse axes are equivalent. Here, we choose the $\hat{y}$-axis as the rotation axis. Figure~\ref{fig:Angle dependence entropy} show the entanglement entropy as a function of the rotation angle $\theta$. Rotating $\ket{m_J=1}$ about the $\hat{y}$-axis by $\theta$ gives,
\begin{multline}
    \hat{R}_y(\theta)\ket{m_J=1}=\frac{1+\operatorname{cos}\theta}{2}\ket{m_J=1}\\
    +\frac{\operatorname{sin}\theta}{\sqrt{2}}\ket{m_J=0}+\frac{1-\operatorname{cos}\theta}{2}\ket{m_J=-1},
\end{multline}
where the corresponding spin vector and the spin tensor are
\begin{equation}
    \vec{S}=(\operatorname{sin}\theta,0,\operatorname{cos}\theta),
\end{equation}
\begin{equation}
    T=\begin{pmatrix}
        \frac{1}{12}-\frac{1}{4}\operatorname{cos}2\theta & 0 & \frac{1}{4}\operatorname{sin}2\theta \\
        0 & -\frac{1}{6} & 0 \\
        \frac{1}{4}\operatorname{sin}2\theta & 0 & \frac{1}{12}+\frac{1}{4}\operatorname{cos}2\theta
    \end{pmatrix}.
\end{equation}
Similarly, rotating $\ket{m_J=0}$ about the $\hat{y}$-axis by $\theta$ gives,
\begin{multline}
    \hat{R}_y(\theta)\ket{m_J=0}=-\frac{\operatorname{sin}\theta}{\sqrt{2}}\ket{m_J=1}\\
    +\operatorname{cos}\theta\ket{m_J=0}+\frac{\operatorname{sin}\theta}{\sqrt{2}}\ket{m_J=-1},
\end{multline}
where the corresponding spin vector and the spin tensor are
\begin{equation}
    \vec{S}=(0,0,0),
\end{equation}
\begin{equation}
    T=\begin{pmatrix}
        -\frac{1}{6}+\frac{1}{2}\operatorname{cos}2\theta & 0 & -\frac{1}{2}\operatorname{sin}2\theta \\
        0 & \frac{1}{3} & 0 \\
        -\frac{1}{2}\operatorname{sin}2\theta & 0 & -\frac{1}{6}-\frac{1}{2}\operatorname{cos}2\theta
    \end{pmatrix}.
\end{equation}
In particular, rotating $\ket{m_J=0}$ and $\ket{m_J=1}$ by $\theta=\pi/2$ corresponds to the maximally and minimally polarized states along the $\hat{x}$-axis, respectively.
As shown in Figure~\ref{fig:Angle dependence entropy}, the entanglement entropy of quarkonium is symmetric about $\theta=\pi/2$, reflecting the symmetry with respect to the transverse plane. For $S$-wave $1^{--}$ quarkonia, the entanglement entropy exhibits only a very weak dependence on $\theta$, indicating an approximately full spatial symmetry. In contrast, for $1^{++},1^{+-}$ and $D$-wave $1^{--}$ quarkonia, the entanglement entropy varies significantly with $\theta$, suggesting that the spatial symmetry is broken and that the longitudinal direction plays a distinguished role. Although the entanglement entropy exhibits a sizable dependence on $\theta$, its variation with the rotation angle remains considerably smaller than the difference between the $\ket{m_J=0}$ and $\ket{m_J=1}$ states. This observation suggests that the dominant polarization dependence of the partonic entanglement entropy arises from the polarization configuration of the meson, while the orientation dependence within a given rotated state gives a subleading effect. Compared with charmonium, bottomonium shows a smaller variation of the entanglement entropy with $\theta$, indicating that heavier quarkonium systems tend to suppress the spatial-symmetry breaking reflected in the entanglement entropy.

\section{Summary and outlook}
\label{sec:Summary and outlook}

In this work, we performed a systematic nonperturbative investigation of quark-antiquark entanglement in heavy quarkonium systems within the light-front Hamiltonian framework. Starting from light-front wave functions (LFWFs), we explicitly constructed the reduced density matrices by tracing over the antiquark degrees of freedom and evaluated the corresponding von Neumann entropies. This formulation establishes a direct connection between quantum-information measures and the transverse-momentum-dependent parton distribution functions (TMDs) of quarkonium. It therefore provides a controlled, QCD-motivated setting for studying how quantum entanglement encodes the nonperturbative dynamics of bound states, extending our previous investigation in a strongly coupled scalar theory in 3+1D~\cite{Zhang:2025ean}.

A central result of this work is that the entanglement entropy is directly linked to the spin-unpolarized and polarized quark TMDs. For spin-0 quarkonia, the quark-antiquark entanglement entropy reduces to the Shannon entropy of the unpolarized TMD $f_1$, up to additive contributions from color and spin degrees of freedom. For spin-1 quarkonia, the entropy acquires a nontrivial dependence on the polarization state, and we derived explicit expressions relating this dependence to polarized and tensor-polarized TMDs. This connection provides a quantum-information perspective on the three-dimensional partonic structure of hadrons and may offer new constraints for phenomenological studies at future electron-ion colliders (EICs).

Using LFWFs obtained within the BLFQ framework, we evaluated the entanglement entropy for charmonium and bottomonium states. The numerical results show that the entropy does not exhibit a clear systematic correlation with the excitation pattern of the state. In particular, for spin-1 quarkonia, the entropy displays a pronounced dependence on the polarization state and rotation angle, revealing a close correlation between entanglement and the spin quantum state of the meson. We also examined the entropy densities in both transverse momentum and longitudinal momentum fraction. The transverse entropy density exhibits peak structures whose multiplicity reflects the excitation pattern of the quarkonium state, while the longitudinal entropy density follows the characteristic behavior of the corresponding parton distribution functions (PDFs).

A key technical component of this work is the treatment of the infrared parameter in the entanglement entropy. Although the momentum-space expression contains an IR-dependent logarithmic contribution, the harmonic oscillator representation provides a finite and discrete formulation once a Gaussian c.m. wave packet is introduced. By matching the momentum-space result to the harmonic-oscillator representation in the narrow-wave-packet limit, we determined the IR cutoff and transverse box size consistently with the BLFQ basis truncation. This matching procedure is essential for obtaining finite and scheme-independent entanglement entropies in the continuum formulation.

The correspondence between entanglement entropy and TMDs established here opens several possible directions for future study. In the valence-dominated regime relevant to heavy mesons and heavy baryons, where contributions from sea quarks and gluons are expected to be suppressed, the entanglement entropy may be indirectly constrained through TMDs measured in semi-inclusive deep inelastic scattering (SIDIS) or Drell-Yan processes~\cite{Bacchetta:2006tn}. For vector mesons, the polarization dependence of the entanglement entropy further suggests that angular correlations and heavy-quarkonium production may provide experimentally accessible quantum-information observables at future colliders~\cite{Dumitru:2023fih,Wu:2024mtj,Sheikh:2024ubi}.

Extending this framework toward full QCD, however, raises several conceptual and technical challenges. First, the inclusion of higher Fock sectors containing sea quarks and gluons generally modifies the direct correspondence between partonic entanglement entropy and quark TMDs. Consequently, for systems that are not dominated by their valence Fock sectors, such as light mesons and baryons, the entanglement entropy cannot in general be reconstructed from quark TMDs alone. And we have to go back to the density matrix constructed from the wave function. Second, the inclusion of dynamical gluons requires Wilson lines in the definition of TMDs to ensure gauge invariance. Correspondingly, a gauge-invariant formulation of the reduced density matrix may require suitable gauge links or gauge-dressed partonic operators. The precise construction, however, depends on the definition of the partonic subsystem and on how the traced and retained degrees of freedom are separated.

A further challenge concerns the resolution and scale dependence of partonic entanglement. In continuum QCD, UV, collinear, and rapidity singularities imply that both the reduced density matrix and its entropy must be defined with respect to a specified renormalization and factorization prescription. The physically relevant subsystem may therefore be a gauge-invariant and experimentally resolvable object, such as a collinear sector or a jet, rather than an isolated bare parton. Although progress has been made in studying the QCD evolution of entanglement entropy in specific kinematic regimes, a general evolution framework consistent with the scale and rapidity evolution of TMDs remains to be established.

Finally, extending the present analysis to more refined quantum-information measures, including the entanglement spectrum~\cite{Li:2008kda}, logarithmic negativity~\cite{Vidal:2002zz}, and entanglement of formation~\cite{Bennett:1996gf}, may reveal features of nonperturbative QCD bound states that are not fully captured by the von Neumann entropy alone. For example, characteristic degeneracies or organizational patterns in the entanglement spectrum may provide information about underlying symmetries of the hadronic wave function. Mixed-state measures become particularly relevant when higher Fock sectors or multiple partonic species are included. In a system containing quarks, antiquarks, and gluons, tracing over all constituents other than a selected pair generally leaves that two-parton subsystem in a mixed state. The von Neumann entropy of either selected parton then reflects both its correlation with the other parton and its correlation with the traced-out environment, and therefore does not by itself quantify the entanglement within the pair. Logarithmic negativity provides a diagnostic of bipartite quantum entanglement based on the partial transpose of the reduced density matrix, whereas the entanglement of formation characterizes the minimum average pure-state entanglement required to represent the mixed two-parton state. Applying these measures to higher Fock sectors would provide a more complete characterization of quantum correlations among quarks, antiquarks, and gluons and would complement the TMD-based description developed in this work.

\vspace{0.5cm}

\section*{Acknowledgements}

This work was supported in part by the Chinese Academy of Sciences under Grant No.~YSBR-101, by the National Natural Science Foundation of China (NSFC) under Grant No.~12375081. 

\appendix
\section{Polarization parameters for spin-1 mesons}\label{Appendix:polarization parameters}
For a spin-1 meson in an arbitrary polarization state, the quantum state can be expanded in terms of the $\hat{J_z}$ eigenstates as,
\begin{multline}
    \ket{\psi(P,\Lambda)}=c_1\ket{\psi(P,m_J=1)}+\\
    c_0\ket{\psi(P,m_J=0)}+c_{-1}\ket{\psi(P,m_J=-1)},
\end{multline}
where the coefficients $c_{1,0,-1}$ satisfy the normalization condition,
\begin{equation}
    |c_1|^2+|c_0|^2+|c_{-1}|^2=1.
\end{equation}
The corresponding polarization parameters of the spin-1 meson can then be expressed in terms of these coefficients as follows,
\begin{equation}
\begin{split}
    &S_T^x=\sqrt{2}\operatorname{Re}(c_1^*c_0+c_0^*c_{-1})\\
    &S_T^y=\sqrt{2}\operatorname{Im}(c_1^*c_0+c_0^*c_{-1})\\
    &S_L=|c_1|^2-|c_{-1}|^2\\
    &S_{TT}^{xx}=-S_{TT}^{yy}=2\operatorname{Re}(c_1^*c_{-1})\\
    &S_{LL}=|c_0|^2-\frac{1}{3}\\
    &S_{TT}^{xy}=2\operatorname{Im}(c_1^*c_{-1})\\
    &S_{LT}^x=\sqrt{2}\operatorname{Re}(c_1^*c_0-c_0^*c_{-1})\\
    &S_{LT}^y=\sqrt{2}\operatorname{Im}(c_1^*c_0-c_0^*c_{-1}).
\end{split}
\end{equation}

\end{document}